\documentclass[a4paper,12pt]{article}
\usepackage{amsmath}
\usepackage{amssymb}
\usepackage[dvips]{graphicx}
\usepackage[dvips]{psfrag}
\usepackage{bm}

\usepackage{wrapfig}

\newcommand{\nn}{\nonumber}
\newcommand{\tr}{\text{tr}\,}

\newcommand{\vev}[1]{\left\langle #1 \right\rangle}

\newcommand{\be}{\begin{equation}}
\newcommand{\ee}{\end{equation}}
\newcommand{\bea}{\begin{eqnarray}}
\newcommand{\eea}{\end{eqnarray}}
\newcommand{\cO}{{\cal O}}
\newcommand{\binomi}[2]{\begin{pmatrix} #1 \\ #2 \end{pmatrix}}

\def\Xint#1{\mathchoice
 {\XXint\displaystyle\textstyle{#1}}%
 {\XXint\textstyle\scriptstyle{#1}}%
 {\XXint\scriptstyle\scriptscriptstyle{#1}}%
 {\XXint\scriptscriptstyle\scriptscriptstyle{#1}}%
 \!\int}
 \def\XXint#1#2#3{{\setbox0=\hbox{$#1{#2#3}{\int}$}
 \vcenter{\hbox{$#2#3$}}\kern-.5\wd0}}
 
 \def\dashint{\Xint-}

\begin{document}
\thispagestyle{empty} \addtocounter{page}{-1}
\begin{flushright}
OIQP-16-05
%
\end{flushright} 
\vspace*{1cm}

\begin{center}
{\large \bf One-point functions of non-SUSY operators at arbitrary genus \\
in a matrix model for type IIA superstrings}\\
\vspace*{2cm}
Tsunehide Kuroki$^*$ and Fumihiko Sugino$^\dagger$\\
\vskip0.7cm
{}$^*${\it General Eduction, National Institute of Technology, Kagawa College}\\
\vspace*{1mm}
{\it 551 Kohda, Takuma-cho, Mitoyo, Kagawa 769-1192, Japan}\\
\vspace*{0.2cm}
{\tt kuroki@dg.kagawa-nct.ac.jp}\\
\vskip0.4cm
{}$^\dagger${\it Okayama Institute for Quantum Physics, } \\
\vspace*{1mm}
{\it Furugyocho 1-7-36, Naka-ku, Okayama 703-8278, Japan}\\
\vspace*{0.2cm}
{\tt fusugino@gmail.com}\\
\end{center}
\vskip1.5cm
\centerline{\bf Abstract}
\vspace*{0.3cm}
{\small 
In the previous paper, the authors pointed out correspondence between 
a supersymmetric double-well matrix model and two-dimensional type IIA superstring theory 
on a Ramond-Ramond background from the viewpoint of symmetry and spectrum. 
This was confirmed by agreement between planar correlation functions in the matrix model 
and tree-level amplitudes in the superstring theory. 
In order to investigate the correspondence further, in this paper we compute
correlation functions to all order of genus expansion in the double scaling limit of the matrix model. 
One-point functions of operators protected by supersymmetry terminate at some finite order, 
whereas those of unprotected operators yield non-Borel summable series. 
The behavior of the latter is characteristic in string perturbation series, providing further evidence 
that the matrix model describes a string theory. 
Moreover, instanton corrections to the planar one-point functions are also computed, and 
universal logarithmic scaling behavior is found for non-supersymmetric operators.

}
\vspace*{1.1cm}



\newpage

\section{Introduction}
The fact that the Large Hadron Collider (LHC) was not able to observe any supersymmetric particles at the first run has made 
supersymmetry breaking predicted to occur at higher energy scale. In this situation, it is meaningful to examine 
a possibility that supersymmetry would be broken in string scale itself rather than lower energy scale. 
On the other hand, by looking back on important roles played by supersymmetry in string theory, we are tempted to 
expect a possibility of spontaneous supersymmetry breaking in string theory. 
In view of these, it is undoubtedly important to explore mechanisms of spontaneous supersymmetry breaking 
in lower-dimensional string theory, namely a toy model of the critical string theory. 

In the previous work \cite{Kuroki:2012nt}, we pointed out 
correspondence between 
a supersymmetric matrix model with the scalar potential 
of double-well type and two-dimensional type IIA superstring theory 
on a Ramond-Ramond background. 
The correspondence has been made based on symmetry and spectrum in both sides. 
In~\cite{Kuroki:2013qpa}, it is explicitly checked by comparing planar correlation functions in the matrix model and those at the tree level in the type IIA superstring theory. 
Based on the correspondence, it is expected that 
the matrix model nonperturbatively realizes the IIA superstring theory by taking its double scaling limit 
and enables nonperturbative investigation for the superstring theory. 
In~\cite{Endres:2013sda,Nishigaki:2014ija}, 
we found that supersymmetries of the matrix model are spontaneously broken due to nonperturbative effects 
and the breaking persists in the double scaling limit, 
which suggests spontaneous breaking of target-space supersymmetries of the IIA theory by its nonperturbative contribution. 

Of course, it is better to make the correspondence firmer by collecting its further evidence, 
in particular by agreement of amplitudes beyond the planar or tree level in both sides. 
In this paper, we present the result of one-point functions in the matrix model to all order of genus expansion 
and their corrections by nonperturbative instanton configurations. 
The forthcoming paper~\cite{paperII} is devoted to the all-order result of two-point 
functions~\footnote
{It seems hard to accomplish direct calculation of multi-point amplitudes in the matrix model 
and worldsheet computation in the IIA superstrings at higher genus.  
One of smarter ways may be to show that the Schwinger-Dyson equations in both sides coincide. 
Then agreement of arbitrary correlation functions at each order in perturbation theory automatically follows 
by matching boundary conditions of these equations. See e.g.\cite{Fukuma:1997en}, 
\cite{Cachazo:2002ry} for such attempts in the context of 
the IIB matrix model \cite{Ishibashi:1996xs}, and the Dijkgraaf-Vafa theory 
\cite{Dijkgraaf:2002dh}, respectively.}. 
In these two papers, we focus on the correlation functions among single-trace operators of a matrix $\phi$: 
\be
\frac{1}{N}\tr\phi^n \qquad (n\in\bm N). 
\label{eq:phi^n}
\ee
For odd $n$, the operator (accompanied with operator mixing) corresponds to a vertex operator of Ramond-Ramond two-form field strength in the IIA theory. 
This is not protected by supersymmetry, while the operator for even $n$ is protected. 
We will see that the correlation functions exhibit totally different behavior depending on odd or even $n$.   

The organization of this paper is as follows. 
We give a brief review of the supersymmetric double-well 
matrix model in the next section. 
Infinitely degenerate vacua appearing in the large-$N$ limit are labeled by filling fractions, and 
we define the partition function with definite filling fraction for finite $N$. 
In section \ref{sec:Nicolai}, we develop a 
general formalism for correlation functions among (\ref{eq:phi^n}) at arbitrary genus for a fixed filling fraction. 
They are obtained from correlation functions of the resolvent operators 
\be
R_2(z)\equiv \frac1N\tr\frac{1}{z-\phi^2}. 
\label{eq:R2op}
\ee   
By the Nicolai mapping, the latter is 
expressed as correlation functions among the resolvent in the Gaussian matrix model at any genus~\footnote{
As discussed in~\cite{Kuroki:2012nt}, 
the Nicolai mapping transforms the operators with even $n$ to single-trace operators of integer power of matrices, 
which are observables in the $c=-2$ topological gravity. 
On the other hand, those with odd $n$ become single-trace operators of half-integer power of 
matrices, which no longer belong to the observables in the topological gravity.}.  
According to this result, 
we explicitly compute the one-point functions in section \ref{sec:one-point}. 
We then confirm the double scaling limit we proposed before indeed works. 
For even $n$, the genus expansion of the one-point functions 
terminates at some finite order and each term does not exhibit any singular behavior 
in the double scaling limit, which is plausible from the viewpoint of supersymmetry-protected operators 
or observables in the $c=-2$ topological gravity. 
On the other hand, for odd $n$, the expansion yields a non-Borel summable series, 
and each term is singular and exhibits universal behavior in the double scaling limit. 
The series grows as $(2h)!$ for large genus $h$, 
which is characteristic behavior in string perturbation series. 
In section \ref{sec:inst}, we consider the planar one-point functions in the presence of instantons 
in the matrix model, namely the leading order of perturbation around nonperturbative objects. 
Subtracting singular nonuniversal parts is necessary in this case differently from the perturbation theory without instanton. 
We see that the subtraction by operator mixing considered in computing cylinder amplitudes 
in~\cite{Kuroki:2012nt} also works here. 
Validity of the double scaling limit is again confirmed. 
In section \ref{sec:fullsector}, we turn to correlation functions evaluated by the total partition function 
in the full sector, namely without specifying the filling fraction. 
By introducing a regularization parameter, we show that possible divergence in the full sector 
can be consistently absorbed by a kind of ``wave-function renormalization'' of odd-power operators 
((\ref{eq:phi^n}) with odd $n$), and thus well-defined correlation functions can be defined. 
Section \ref{sec:discussion} is devoted to discussions. 
In appendix~\ref{app:recursion}, we solve a recursion relation for coefficients in genus expansion of 
the resolvent in the Gaussian matrix model. 
In appendix \ref{app:evenone-pt}, we compute the one-point functions 
of even-power operators ((\ref{eq:phi^n}) with even $n$) in a more general setting than the text. 
In appendix~\ref{app:distortion}, the instanton effect in section~\ref{sec:inst} is reproduced 
from the viewpoint of distortion of the eigenvalue distribution by the instantons. 

\section{Review of the supersymmetric matrix model}
\label{sec:MM_review}
\setcounter{equation}{0}
We consider a supersymmetric matrix model defined by the action: 
\begin{align}
S = N \tr \left[\frac12 B^2 +iB(\phi^2-\mu^2) +\bar\psi (\phi\psi+\psi\phi)\right],  
\label{eq:S}
\end{align}
where $B$, $\phi$ are Grassmann even, and $\psi$, $\bar\psi$ are Grassmann odd 
$N\times N$ Hermitian matrices, respectively. 
By completing the square with respect to $B$, we find that the scalar potential for $\phi$ is of 
double-well type. 
The action $S$ is invariant under supersymmetry transformations generated by $Q$ and $\bar{Q}$: 
\begin{align}
Q\phi =\psi, \quad Q\psi=0, \quad Q\bar{\psi} =-iB, \quad QB=0, 
\label{eq:QSUSY}
\end{align}
and 
\begin{align}
\bar{Q} \phi = -\bar{\psi}, \quad \bar{Q}\bar{\psi} = 0, \quad 
\bar{Q} \psi = -iB, \quad \bar{Q} B = 0,  
\label{eq:QbarSUSY}
\end{align}
which lead to the nilpotency: $Q^2=\bar{Q}^2=\{ Q,\bar{Q}\}=0$. 

As shown in \cite{Kuroki:2009yg, Kuroki:2010au}, 
the planar limit of the matrix model for $\mu^2\geq 2$ has infinitely degenerate supersymmetric vacua 
parametrized by filling fractions $(\nu_+, \nu_-)$, 
which represent configurations that $\nu_\pm N$ of the eigenvalues of $\phi$ are around the minimum $x=\pm |\mu |$  of the double-well potential $\frac12(x^2-\mu^2)^2$. On the other hand, that for $\mu^2<2$ has a vacuum which breaks the supersymmetry. 
The boundary $\mu^2=2$ is a critical point at which the third-order phase transition occurs. 
A simple large-$N$ limit (planar limit) remains only the planar diagrams (tree amplitudes in the corresponding string theory). 
In fact, we have explicitly seen in~\cite{Kuroki:2013qpa} that the result of several types of correlation functions 
in the matrix model~\cite{Kuroki:2012nt} reproduces 
the tree amplitudes in two-dimensional type IIA superstring theory on a nontrivial Ramond-Ramond background.  
As a limit yielding amplitudes beyond the planar ones, we consider the following double scaling limit~\cite{Endres:2013sda} that 
approaches the critical point from the inside of the supersymmetric phase: 
\begin{align}
N\rightarrow\infty, \quad \mu^2\rightarrow 2+0, \quad  \mbox{with} \quad  
s=N^{\frac23}(\mu^2-2):\mbox{~fixed}.
\label{eq:dsl}
\end{align}
This limit of the matrix model is expected to provide nonperturbative formulation of the superstring theory with string coupling constant $g_s$ proportional to $s^{-\frac32}$. 
In~\cite{Endres:2013sda,Nishigaki:2014ija}, instanton contribution to the free energy 
of the matrix model is found to have a factor $\exp\left(-\frac{C}{g_s}\right)$ 
with a constant $C$ of $\cO(1)$. 
This form is typical of solitonic objects in string theory (D-branes). 
Furthermore, the instantons cause spontaneous supersymmetry breaking in the matrix model, 
which implies violation of target-space supersymmetry 
by nonperturbative effects in the corresponding superstring theory.   

In this paper we are interested in correlation functions at higher genera 
in the double scaling limit \eqref{eq:dsl} in each sector with a fixed filling fraction. 
More precisely, in terms of the eigenvalues of $\phi$, the partition function of \eqref{eq:S} is given as 
\cite{Kuroki:2012nt,Endres:2013sda}
\bea
Z& \equiv&(-1)^{N^2}\int d^{N^2}B\,d^{N^2}\phi\,\left(d^{N^2}\psi\,d^{N^2}\bar{\psi}\right)\,e^{-S} \nn \\
&=&\tilde C_N\int_{-\infty}^{\infty}
\left(\prod_{i=1}^N2\lambda_id\lambda_i\right)\triangle(\lambda^2)^2\,
e^{-N\sum_{i=1}^N\frac12(\lambda_i^2-\mu^2)^2}, 
\label{eq:Z}
\eea
where the normalization of the measure is fixed by 
\be
\int d^{N^2}\phi\,e^{-N\tr(\frac12\phi^2)}=\int d^{N^2}B\,e^{-N\tr(\frac12B^2)}=1
\label{eq:measure1}
\ee 
and 
\be
(-1)^{N^2}\int\left(d^{N^2}\psi \,d^{N^2}\bar{\psi}\right) e^{-N\tr(\bar{\psi}\psi)}=1. 
\label{eq:measure2}
\ee
$\tilde C_N$ is a constant dependent only on $N$: 
$\tilde{C}_N=(2\pi)^{-\frac{N}{2}}N^{\frac{N^2}{2}}\left(\prod_{k=0}^Nk!\right)^{-1}$~\cite{Kuroki:2010au}, and 
$\triangle(x)$ stands for the Vandermonde determinant for eigenvalues $x_i$ ($i=1,\cdots, N$): $\triangle(x) \equiv \prod_{i>j}(x_i-x_j)$. 
Namely, $\triangle(\lambda^2)= \prod_{i>j}(\lambda_i^2-\lambda_j^2)$. 
By dividing the integration region of each $\lambda_i$ according to the filling fraction, 
the total partition function can be expressed as a sum of 
each partition function with a fixed filling fraction: 
\begin{align}
&Z=\sum_{\nu_-N=0}^{N}\frac{N!}{(\nu_+N)!(\nu_-N)!}\,Z_{(\nu_+,\nu_-)}, \nn \\ 
&Z_{(\nu_+,\nu_-)}\equiv \tilde C_N
\int_0^{\infty}\left(\prod_{i=1}^{\nu_+N}2\lambda_id\lambda_i\right)
\int_{-\infty}^0\left(\prod_{j=\nu_+N+1}^N2\lambda_jd\lambda_j\right)
\triangle(\lambda^2)^2 
\,e^{-N\sum_{m=1}^N\frac12(\lambda_m^2-\mu^2)^2}. 
\end{align}
By changing the integration variables $\lambda_j\rightarrow -\lambda_j$ ($j=\nu_+N+1,\cdots, N$), it is easy to find 
\begin{align}
Z_{(\nu_+, \nu_-)}=(-1)^{\nu_-N}Z_{(1,0)}, 
\label{eq:Z_MM}
\end{align}
and therefore the total partition function vanishes:~\footnote{
The consequence directly follows from the fact that the integrand of the total partition function (\ref{eq:Z}) is odd under the sign flip 
of an arbitrary eigenvalue.} 
\begin{align}
Z=\sum_{\nu_-N=0}^{N}\frac{N!}{(\nu_+N)!(\nu_-N)!}(-1)^{\nu_-N}Z_{(1,0)} = (1+(-1))^N Z_{(1,0)}=0. 
\label{eq:totalZ}
\end{align}

We define the correlation function of $K$ single-trace operators $\frac{1}{N}\tr \cO_a(\phi)$ ($a=1,\cdots, K$) in the $(\nu_+,\nu_-)$ sector as 
\begin{align}
\vev{\prod_{a=1}^K\frac{1}{N}\tr{\cal O}_a(\phi)}^{(\nu_+,\nu_-)} &\equiv \frac{\tilde C_N}{Z_{(\nu_+,\nu_-)}}
\int_0^{\infty}\left(\prod_{i=1}^{\nu_+N}2\lambda_id\lambda_i\right)
\int_{-\infty}^0\left(\prod_{j=\nu_+N+1}^N2\lambda_jd\lambda_j\right)\triangle(\lambda^2)^2\nn \\
& \hspace{20mm}\times \left(\prod_{a=1}^K\frac{1}{N}\sum_{i=1}^N{\cal O}_a(\lambda_i)\right)\,
e^{-N\sum_{m=1}^N\frac12(\lambda_m^2-\mu^2)^2},  
\label{eq:correlator}
\end{align}
and express its connected part by the $1/N$-expansion:  
\begin{align}
\vev{\prod_{a=1}^K \frac{1}{N}\tr{\cal O}_a(\phi)}_C^{(\nu_+,\nu_-)}=\sum_{h=0}^\infty \frac{1}{N^{2h+2K-2}}\,
\vev{\prod_{a=1}^K \frac{1}{N}\tr{\cal O}_a(\phi)}_{C,\,h}^{(\nu_+,\nu_-)}.  
\label{eq:vev_MM}
\end{align}
$\vev{\,\cdot\,}_{C,\,h}^{(\nu_+,\nu_-)}$ denotes the connected correlation function on a handle-$h$ random surface 
with the $N$-dependence factored out; i.e., the quantity of ${\cal O}(N^0)$. 
In this paper, we focus on the case that $\cO_a(\phi)$ are polynomials of $\phi$. 
Operators $\frac{1}{N}\tr B^k$ or equivalently (linear combinations of) 
$\frac{1}{N}\tr\phi^{2k}$ ($k\in\bm N\cup\{0\}$) are invariant 
under the supersymmetries (\ref{eq:QSUSY}) and (\ref{eq:QbarSUSY}). 
Correlation functions among them do not exhibit any nonanalytic behavior as $s\to 0$ 
at the planar level ($h=0$)~\cite{Kuroki:2012nt}, which is characteristic of protection by supersymmetry.  
On the other hand, operators of odd powers: $\frac{1}{N}\tr \,\phi^{2k+1}$ ($k\in\bm N\cup\{0\}$) 
are not invariant under either of $Q$ or $\bar{Q}$, and 
show nontrivial critical behavior as power of $\ln s$ at the planar level \cite{Kuroki:2012nt}: 
\bea
\left.\vev{\prod_{a=1}^K\Phi_{2k_a+1}}_{C,\,0}^{(\nu_+,\nu_-)}\right|_{\rm sing.} & = &
(\nu_+-\nu_-)^K\,(\mbox{const.})\,s^{2-\gamma+\sum_{a=1}^K(k_a-1)}\,(\ln s)^K \nn \\
& &+(\mbox{less singular at $s=0$}) 
\label{eq:higher_phi}
\eea
with the string susceptibility exponent $\gamma=-1$. 
Here, ``$|_{\rm sing.}$'' stands for ignoring regular functions of $s$ at $s=0$, and 
the factor ``(const.)'' contains a certain power of $N$. 
$\Phi_{2k+1}$ is essentially $\frac1N\tr\phi^{2k+1}$ with operator mixing: 
\begin{align}
\Phi_{2k+1}=\frac{1}{N}\,\tr\phi^{2k+1} + (\mbox{mixing}) \qquad (k\in\bm N\cup\{0\}),  
\label{eq:mixing}
\end{align}
where ``mixing'' represents a sum of even powers of $\phi$ lower than the degree $2k+1$.  
For instance, an explicit form is given in  \eqref{Phi-mixing}. 
These are introduced in order to remove nonuniversal singular terms in the double scaling limit.
\eqref{eq:higher_phi} are expected to correspond to correlation functions 
of the Ramond-Ramond fields in the two-dimensional type IIA superstring theory~\cite{Kuroki:2012nt}.    
By computing scattering amplitudes in the superstring side and comparing the result with (\ref{eq:higher_phi}), 
the expectation has been confirmed for one- and two-point functions 
($K=1,\,2$) with arbitrary odd powers~\cite{Kuroki:2013qpa}. 
Thus it is desirable to go beyond the tree level and get an expression as in \eqref{eq:higher_phi} 
at higher genus, which is our main motivation.

\section{Correlation functions at arbitrary genus}
\label{sec:Nicolai}
\begin{figure}[h]
\centering
\includegraphics[width=8cm, bb=0 330 700 720, clip]{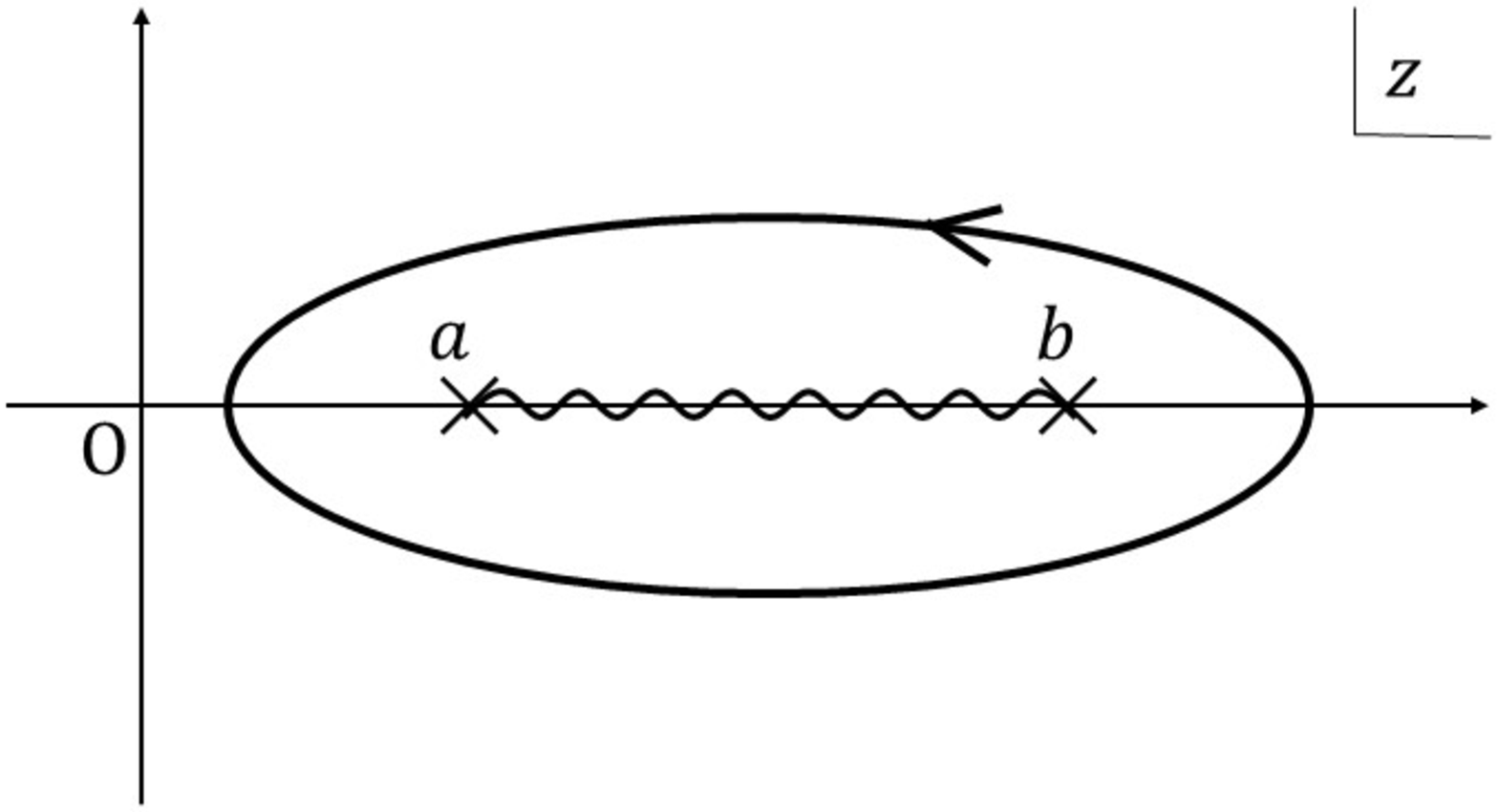}
\caption{\small Integration contour on the complex $z$-plane.}
\label{fig:zcontour}
\vspace{7mm}
\end{figure}

\setcounter{equation}{0}
In order to compute one-point functions for a function ${\cal O}(\phi)$ of $\phi$  
in the filling fraction $(1,0)$ at arbitrary genus $h$: $\vev{\frac{1}{N}\tr{\cal O}(\phi)}_h^{(1,0)}$, 
we consider the $\phi^2$-resolvent 
\begin{align}
\vev{R_2(z^2)}_{h}^{(1,0)}=\vev{\frac1N\tr\frac{1}{z^2-\phi^2}}_{h}^{(1,0)}  
\label{eq:R2}
\end{align}
rather than the standard resolvent $\vev{\frac1N\tr\frac{1}{z-\phi}}_h^{(1,0)}$. 
The former is protected by the supersymmetry and is expected to have a simpler form compared with the latter. 
In terms of the eigenvalues, $R_2(z^2)$ is  
\be
\frac1N\sum_{i=1}^N \frac{1}{z^2-\lambda_i^2}= \frac1N\frac{1}{2z}\sum_{i=1}^N\left(\frac{1}{z-\lambda_i}+\frac{1}{z+\lambda_i}\right)
\ee
and $\frac{1}{z-\lambda_i}$ ($\frac{1}{z+\lambda_i}$) has poles 
only on the positive (negative) real axis in the $(1,0)$ filling fraction. 
At each order in the $1/N$-expansion, the poles accumulate to develop a cut $[a,b]$ ($[-b,-a]$), 
where $a=\sqrt{\mu^2-2}$ and $b=\sqrt{\mu^2+2}$~\cite{Kuroki:2012nt}. 
Thus the one-point function is given by the contour integral of (\ref{eq:R2}):\footnote
{Here we implicitly assume that ${\cal O}(z)$ itself does not have singularity on the cut $[a,b]$.} 
\be
\vev{\frac{1}{N}\tr{\cal O}(\phi)}_h^{(1,0)}= \oint_{[a,b],\,z} 2z\cdot {\cal O}(z)
\vev{R_2(z^2)}_{h}^{(1,0)},   
\label{eq:1pt1}
\ee
where $\oint_{D,~z}\equiv \oint_{D}\frac{dz}{2\pi i}$ denotes the $z$-integral 
along the contour encircling only the region $D$ counterclockwise~\footnote{
In this paper, we treat cases where $D$ is an interval or a point. When $D$ is their union, the contour 
should be understood as $\oint_{D_1\cup D_2,~z}=\oint_{D_1,~z}+\oint_{D_2,~z}$.
\label{footnote:union}}.  
The case of $D=[a,\,b]$ is depicted in Fig.~\ref{fig:zcontour}. 
In the case of ${\cal O}(\phi)=\phi^n$ as in \eqref{eq:phi^n}, the one-point function for a general filling fraction $\vev{\frac{1}{N}\tr\phi^n}_h^{(\nu_+,\,\nu_-)}$ is obtained by simply multiplying 
the factor $(\nu_+-\nu_-)^\sharp$ to (\ref{eq:1pt1}), where $\sharp=0$ and $1$ for even and odd  $n$, respectively~\cite{Kuroki:2012nt}. 

It is easy to extend this argument to multi-point functions. The $K$-point connected correlation function among $\frac{1}{N}\tr{\cal O}_{a}(\phi)$ 
($a=1,\cdots, K$) is obtained from the $K$-point function of the $\phi^2$-resolvent as 
\be
\vev{\prod_{\ell=1}^K\frac{1}{N}\tr{\cal O}_\ell(\phi)}_{C,\,h}^{(1,0)}
= \prod_{\ell=1}^K\oint_{[a,b],\,z_\ell} 2z_\ell\cdot {\cal O}_{\ell}(z_{\ell})
\vev{\prod_{a=1}^KR_2(z_a^2)}_{C,\,h}^{(1,0)}. 
\label{eq:Kpt1}  
\ee

Let us next note that the $\phi^2$-resolvent is mapped to the resolvent in a Gaussian matrix model. 
In fact, the Nicolai mapping $x_i=\mu^2-\lambda_i^2$ ($i=1,\cdots, N$) recasts 
the partition function $Z_{(1,0)}$ and the correlation function (\ref{eq:correlator}) 
in the $(1,0)$ sector as 
\be
Z_{(1,0)}= \tilde{C}_N\int_{-\infty}^{\mu^2}\left(\prod_{i=1}^Ndx_i\right) \triangle(x)^2\,e^{-N\sum_{i=1}^N\frac12x_i^2} \,\equiv Z^{(G')} 
\label{eq:Z(1,0)_Nicolai} 
\ee
and 
\bea
& & \vev{\prod_{a=1}^K\frac{1}{N}\tr{\cal O}_a(\phi)}^{(1,0)} = \frac{\tilde C_N}{Z^{(G')}}
\int_{-\infty}^{\mu^2}\left(\prod_{i=1}^Ndx_i\right)\triangle(x)^2
\left(\prod_{a=1}^K\frac{1}{N}\sum_{i=1}^N{\cal O}_a\left(\sqrt{\mu^2-x_i}\right)\right)\nn \\
& & \hspace{60mm}\times e^{-N\sum_{i=1}^N\frac12x_i^2},  
\label{eq:correlator_Nicolai}
\eea
respectively. 
The integrals of the eigenvalues $x_i$ are not over the entire real line, but are bounded 
from above by $\mu^2$. 
We put the superscript `$(G')$' on quantities in such a Gaussian matrix model. 
The difference from the standard one whose eigenvalues are integrated over the whole real axis is nonperturbative in $1/N$ and negligible in the genus expansion~\cite{Endres:2013sda}. 
As pointed out in~\cite{Kuroki:2012nt}, supersymmetric operators $\frac{1}{N}\tr\phi^{2k}$ are mapped onto observables in 
the $c=-2$ topological gravity (the standard Gaussian matrix model); i.e., polynomials in $x_i$, while non-supersymmetric operators 
$\frac{1}{N} \tr\phi^{2k+1}$ are not due to the branch cut singularity of the square root. 
In particular, $R_2(z^2)$ becomes 
\be
\frac{1}{N}\sum_{i=1}^N\frac{1}{z^2-\mu^2+x_i}\,= -R_M(\mu^2-z^2)
\ee
in terms of eigenvalues, 
where $R_M(x)\equiv \frac{1}{N}\tr\frac{1}{x-M}$ is the resolvent in the Gaussian matrix model whose matrix variable $M$ is an $N\times N$ 
Hermitian matrix and its eigenvalues are $x_i$ ($i=1,\cdots,N$). 
Now the problem is reduced to higher-genus correlation functions in the Gaussian matrix model: 
\be
\vev{\frac{1}{N}\tr{\cal O}(\phi)}_h^{(1,0)}=-\oint_{[a,b],\,z}2z\cdot {\cal O}(z)
\vev{R_M(\mu^2-z^2)}_h^{(G)} \label{eq:1pt2} 
\ee
and 
\be
\vev{\prod_{a=1}^K\frac{1}{N}\tr{\cal O}_a(\phi)}_{C,\,h}^{(1,0)}
= \prod_{\ell=1}^K\left(-\oint_{[a,b],\,z_\ell} 2z_\ell\cdot {\cal O}_\ell(z_{\ell})\right)
\vev{\prod_{a=1}^KR_M(\mu^2-z_a^2)}_{C,\,h}^{(G)}.  
\label{eq:Kpt} 
\ee
The superscript `$(G)$' (not `$(G')$') on the r.h.s. indicates the use of the standard Gaussian matrix model, 
which is allowed in the genus expansion from the reason mentioned above. In particular, 
\begin{align}
\vev{\prod_{a=1}^K\frac{1}{N}\tr\phi^{n_a}}_{C,\,h}^{(1,0)}
= \prod_{\ell=1}^K\left(-\oint_{[a,b],\,z_\ell} 2z_\ell\cdot z_{\ell}^{n_{\ell}}\right)
\vev{\prod_{a=1}^KR_M(\mu^2-z_a^2)}_{C,\,h}^{(G)}.  
\label{eq:Kpt_power} 
\end{align}

It is interesting to point out that correlation functions involving not only even-power operators but also odd-power ones 
are obtained from the supersymmetric $\phi^2$-resolvent as in (\ref{eq:Kpt_power}). 
It suggests that infinitely many supersymmetric local operators ($\frac{1}{N}\tr\phi^{2k}$ 
($k\in\bm N$)) that are equivalent to the resolvent operator (\ref{eq:R2op}) 
can also carry information for non-supersymmetric operators.

\section{One-point functions}
\label{sec:one-point}
\setcounter{equation}{0}
In this section, we give explicit formulas of the one-point functions 
of the operators (\ref{eq:phi^n}) at arbitrary genus from (\ref{eq:Kpt_power}). 

According to the literature in the random matrix theory, e.g.~\cite{HT_2012}, 
the $1/N$-expansion of the resolvent in the Gaussian matrix model is explicitly given as 
\be
\vev{R_M(x)}^{(G)}=\sum_{h=0}^{\infty}\frac{1}{N^{2h}}\eta_h(x)
\ee
with 
\begin{align}
&\eta_0(x)=\frac12x-\frac12(x^2-4)^{\frac12}, 
\label{eq:eta0} \\
&\eta_j(x)=\sum_{r=2j}^{3j-1}C_{j,\,r}(x^2-4)^{-r-\frac12} \quad (j\in\bm N).
\label{eq:etah}
\end{align}
The coefficients $C_{j,\,r}$ satisfy a recursion relation 
\begin{align}
C_{j+1,\,r}=\frac{(2r-3)(2r-1)}{r+1}\left((r-1)C_{j,\,r-2}+(4r-10)C_{j,\,r-3}\right) 
\label{eq:recursion}
\end{align}
for $2j+2\leq r\leq 3j+2$ with conditions 
\begin{align}
C_{j,\,2j-1}=C_{j,\,3j}=0, \qquad C_{1,\,2}=1.
\label{eq:Cbdy}
\end{align} 
The use of this result in (\ref{eq:Kpt_power}) with $x=\mu^2-z_1^2$ leads to
\bea
& & \vev{\frac{1}{N}\tr\phi^{n}}^{(1,0)}_0 = -\frac12I_{1,\,n}, \nn \\
& & \vev{\frac{1}{N}\tr\phi^{n}}^{(1,0)}_j = \sum_{r=2j}^{3j-1}C_{j,\,r}\,I_{-2r-1,\,n}\quad (j\in\bm N), 
\label{eq:1pt3}
\eea
where
\be
I_{m,n}\equiv \oint_{[-2,\,2],\,x}(x^2-4)^{\frac{m}{2}}(\mu^2-x)^{\frac{n}{2}}. 
\label{eq:Imn}
\ee

\subsection{Computation of universal contribution}
\label{subsec:ACbyBeta}
We evaluate universal contribution to the one-point functions (\ref{eq:1pt3}) in the double scaling limit~(\ref{eq:dsl}). 
When $n$ is an even integer, $I_{m,\,n}$ becomes a polynomial of $\mu^2$ and the one-point functions do not exhibit any nonanalytic 
behavior as $\mu^2\to 2$. Therefore, we focus on the case where $n$ is odd 
($n=2k+1$, $k\in\bm N\cup\{0\}$) in this subsection. 
Then the mixing term in (\ref{eq:mixing}) solely gives uninteresting analytic contribution 
to the one-point functions, and  
$\vev{\Phi_{2k+1}}^{(1,0)}_h$ is identical with $\vev{\frac{1}{N}\tr\phi^{2k+1}}^{(1,0)}_h$ regarding their universal (nonanalytic) contribution. 

\begin{figure}[h]
\centering
\includegraphics[width=8cm, bb=0 330 720 710, clip]{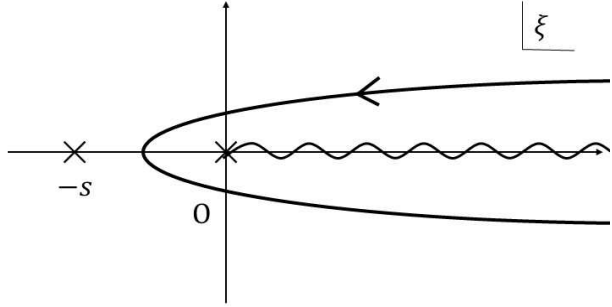}
\caption{Integration contour on the complex $\xi$-plane.}
\label{fig:xicontour}
\vspace{7mm}
\end{figure}

Let us change variables so that they will magnify the vicinity of the critical point in the double scaling limit 
as $\mu^2=2+N^{-\frac23}s$, $x=2-N^{-\frac23}\xi$. Then 
\begin{align}
I_{m,\,2k+1}=-\left(N^{-\frac23}\right)^{\frac{m+3}{2}+k}(-2i)^m
\oint_{[0,\infty),\,\xi}\xi^{\frac m2}(s+\xi)^{k+\frac12}
\left(1+{\cal O}(N^{-\frac23})\right)  
\label{eq:Ibyxi}
\end{align}
with $m=1,\,-1,\,-3,\cdots$ and $k=0,\,1,\,2,\cdots$. 
Since $x=2$ and $x=-2$ are mapped to $\xi=0$ and $\xi=4N^{\frac23}$ respectively, 
the integration contour becomes to surround the positive real axis in the double scaling limit 
as in Fig. \ref{fig:xicontour}. 

The integration around the critical point $\xi=0$ provides the universal contribution, while possible divergence 
from the integration around $\xi= \infty$ will become cut-off dependent ($N$-dependent) analytic terms of $s$ (nonuniversal 
contribution). The latter is discarded by taking $s$-derivatives $(k+3)$ times:    
\be
\frac{\partial^{k+3}}{\partial s^{k+3}}I_{m,\,2k+1}= -\left(N^{-\frac23}\right)^{\frac{m+3}{2}+k}(-2i)^m
\frac{\Gamma(k+\frac32)}{\Gamma(-\frac32)}\oint_{[0,\infty),\,\xi}\xi^{\frac m2}(s+\xi)^{-\frac52}\left(1+{\cal O}(N^{-\frac23})\right).  
\ee
By rescaling $\xi$ as $\xi\to s\xi$, the integral is expressed by the Beta function as~\footnote{
Note that the expression of the Beta function 
$B(p,\,q)=\frac{2\pi i}{e^{i2\pi p}-1}\oint_{[0,\infty),\,y}\frac{y^{p-1}}{(1+y)^{p+q}}$ is valid for a larger region 
$p\notin \bm Z$ and $\mbox{Re }q>0$ compared with $\mbox{Re }p>0$ and $\mbox{Re }q>0$ 
where another integral representation 
$B(p,\,q) = \int_0^\infty dy\frac{y^{p-1}}{(1+y)^{p+q}}$ is available. 
Furthermore, the integral $\oint_{[0,\infty),\,y}\frac{y^{p-1}}{(1+y)^{p+q}}$ itself 
is well-defined for $p\in {\bf C}$ and $\mbox{Re }q>0$. The combination $(e^{i2\pi p}-1)\Gamma(p)$ has no singularity 
except for $p=\infty$.}  
\be
\oint_{[0,\infty),\,\xi}\xi^{\frac m2}(1+\xi)^{-\frac52}= \frac{e^{i\pi m}-1}{2\pi i}B\left(\frac{m}{2}+1,\,\frac{3-m}{2}\right). 
\ee
Thus we have 
\begin{align}
\frac{\partial^{k+3}}{\partial s^{k+3}}I_{m,\,2k+1}= \left(N^{-\frac23}\right)^{\frac{m+3}{2}+k}(-i)^{m+1}\frac{2^m}{\pi^2}
\Gamma\left(k+\frac32\right)\Gamma\left(\frac{m}{2}+1\right)\Gamma\left(\frac{3-m}{2}\right)\,s^{-\frac{3-m}{2}}
\label{eq:der_I}
\end{align}
up to subleading contribution. 
Integrating $(k+3)$ times with respect to $s$ leads to  
\be
I_{m,\,2k+1}=
-\left(N^{-\frac23}\right)^{\frac{m+3}{2}+k}\frac{2^m}{\pi^2}
\frac{\Gamma\left(\frac{m}{2}+1\right)\Gamma\left(k+\frac32\right)}
{\Gamma\left(\frac{m+5}{2}+k\right)}
\,s^{\frac{m+3}{2}+k}\ln s +(\mbox{less singular})
\label{eq:Iresult1}
\ee
for $\frac{m+3}{2}+k\geq 0$, and 
\begin{align}
I_{m,\,2k+1}=&
\left(N^{-\frac23}\right)^{\frac{m+3}{2}+k}(-1)^{\frac{m+3}{2}+k}\frac{2^m}{\pi^2}
\Gamma\left(\frac{m}{2}+1\right)\Gamma\left(k+\frac32\right)\Gamma\left(-\frac{m+3}{2}-k\right)
s^{\frac{m+3}{2}+k} \nn \\
&+(\mbox{less singular})
\label{eq:Iresult2}
\end{align}
for $\frac{m+3}{2}+k< 0$. 
Here, ``(less singular)'' stands for less singular terms at $s=0$ compared with the first term. 
More precisely, they are polynomials in $\mu^2=2+N^{-\frac23}s$ 
(cutoff dependent nonuniversal parts) or 
subleading terms in the double scaling limit~(\ref{eq:dsl}). 
Since from the definition~(\ref{eq:Imn}), $I_{m,\,n}$ depends on $s$ and $N$ only through the combination $N^{-\frac23}s$, 
the r.h.s. of (\ref{eq:Iresult1}) should appear as 
\be
\left(N^{-\frac23}s\right)^{\frac{m+3}{2}+k}\ln\left(N^{-\frac23}s\right) 
= \left(N^{-\frac23}s\right)^{\frac{m+3}{2}+k}\ln s 
-\frac23 \left(N^{-\frac23}s\right)^{\frac{m+3}{2}+k}\ln N
\label{eq:lnN}
\ee
up to the overall factor independent of $s$ and $N$. 
Note that although the last term is larger than the first term in the double scaling limit, 
it belongs to the less singular terms around $s=0$. 

{}From~\eqref{eq:Iresult1} and \eqref{eq:Iresult2}, we see that the universal part of $I_{m,n}$ is more dominant 
for smaller $m$ with $k$ fixed. Hence in the sum in \eqref{eq:1pt3}, 
$r=3j-1$ gives the most dominant contribution: 
\begin{align}
\vev{\frac1N\tr\phi^{2k+1}}_j^{(1,0)}
=C_{j,\,3j-1}I_{-6j+1,\,2k+1}
\label{eq:one-pttemp}
\end{align}
in the double scaling limit. 
As in appendix~\ref{app:recursion}, the coefficient $C_{j,\,3j-1}$ can be solved in a simple form~(\ref{C_r=3j-1}). 
It can be rewritten as   
\begin{align}
C_{j,\,3j-1}=\frac{1}{4\sqrt{\pi}}\left(\frac{16}{3}\right)^j\frac{\Gamma\left(3j-\frac12\right)}{j!}. 
\label{eq:Crelevant}
\end{align}
We thus find 
the relevant contribution to the one-point function \eqref{eq:1pt3} 
in the double scaling limit as 
\be
\left.\vev{\frac1N\tr\phi^{2k+1}}_{h}^{(1,0)}\right|_{\text{univ.}} 
=(N^{-\frac23})^{k+2-3h}\frac{1}{2\pi^{\frac32}}\frac{1}{h!}
\left(-\frac{1}{12}\right)^h
\frac{\Gamma\left(k+\frac32\right)}{\Gamma\left(k+3-3h\right)}\,s^{k+2-3h}\ln s
\label{eq:1pt_h1}
\ee
for $0\leq h\leq \frac{k+2}{3}$, and  
\be
\left.\vev{\frac1N\tr\phi^{2k+1}}_{h}^{(1,0)}\right|_{\text{univ.}} =
(N^{-\frac23})^{k+2-3h}\frac{(-1)^{k+1}}{2\pi^{\frac32}}\frac{1}{h!}
\left(\frac{1}{12}\right)^h
\Gamma\left(k+\frac32\right)\Gamma\left(3h-k-2\right)\,s^{k+2-3h}
\label{eq:1pt_h2}
\ee
for $h>\frac{k+2}{3}$. 
The symbol ``$|_{\rm univ.}$'' means the most dominant nonanalytic term at $s=0$ (the universal part) 
taken~\footnote{The $h=0$ case in (\ref{eq:1pt_h1}) reproduces the result given in \cite{Kuroki:2012nt} with $\omega = N^{-\frac23}\frac{s}{4}$.}. 
In other words, recalling \eqref{eq:vev_MM}, we obtain the genus expansion of the universal part 
of the one-point function as 
\begin{align}
&\left.\vev{\frac1N\tr\phi^{2k+1}}^{(1,0)}\right|_{\text{univ.,\,pert.}} \nn \\
&=N^{-\frac23(k+2)}\frac{\Gamma\left(k+\frac32\right)}{2\pi^{\frac32}} 
\biggl[\sum_{h=0}^{\left[\frac13(k+2)\right]}\frac{1}{h!}\left(-\frac{1}{12}\right)^h
\frac{1}{\Gamma\left(k+3-3h\right)}s^{k+2-3h}\ln s \nn \\
&\phantom{N^{-\frac23(k+2)}}
+(-1)^{k+1}\sum_{h=\left[\frac13(k+2)\right]+1}^{\infty}\frac{1}{h!}\left(\frac{1}{12}\right)^h
\Gamma\left(3h-k-2\right)s^{k+2-3h}\biggr].
\label{eq:one-ptgenusexp}
\end{align}
Here, the subscript ``pert." indicates the genus expansion (the expansion with respect to 
$s^{-3}\propto N^{-2}\propto g_s^2$) 
that corresponds to perturbative expansion in string theory~\footnote{
Here we do not take into account nonperturbative contribution from the boundary of the eigenvalue integration as 
mentioned in section~\ref{sec:Nicolai}.}.  
$\left[\frac13(k+2)\right]$ denotes the greatest integer not exceeding $\frac13(k+2)$. 
The result~(\ref{eq:one-ptgenusexp}) explicitly shows that 
the double scaling limit (\ref{eq:dsl}) keeps contribution of each order in the genus expansion finite. 
The overall factor $N^{-\frac23(k+2)}$ can be absorbed in the ``wave function 
renormalization'' of the operator $\frac1N\tr\phi^{2k+1}$.

The logarithmic singularity appears only at lower genera. This reminds us of the case of 
the bosonic $c=1$ noncritical string theory on two-dimensional target space. There, the free energy has the logarithmic behavior 
only at genus zero and one \cite{Gross:1990ub,Distler:1990mt}. 

We can read large-order behavior from \eqref{eq:one-ptgenusexp}. In fact, relevant 
higher-genus contribution is given in the second line on the r.h.s.. 
This is a positive term series growing as $(2h)!$:  
\be
(2h)!\left(\frac43 s^{\frac32}\right)^{-2h}
\label{eq:large-order}
\ee
for sufficiently large genus $h$.  
It is in contrast to the lower-genus contribution 
given in the first line that is an alternating series. 
The behavior (\ref{eq:large-order}) is a characteristic feature of string perturbation series, 
and gives a further support that the matrix model describes a string theory 
in the double scaling limit \cite{Shenker:1990uf}. 
As discussed in~\cite{Kuroki:2012nt}, the matrix model with two supersymmetries~\eqref{eq:QSUSY} and \eqref{eq:QbarSUSY} can be regarded as 
two-dimensional type IIA superstring theory with the corresponding target-space supersymmetries.  
In the matrix model, the supersymmetries are preserved to all order in the $1/N$-expansion, 
while they are spontaneously broken due to nonperturbative effects~\cite{Endres:2013sda,Nishigaki:2014ija}. 
This indicates that the supersymmetries in the IIA superstring theory are not broken to all order in perturbation theory 
(expansion in $s^{-3}(=g_s^2)$), but are violated nonperturbatively.   

Furthermore, as in \cite{Shenker:1990uf}, from (\ref{eq:large-order}) 
we can also deduce a nonperturbative effect as 
$\exp\left(-\frac43 s^{\frac32}\right)\left(=\exp\left(-\frac43\frac{1}{g_s}\right)\right)$, 
which in fact coincides with one calculated in \cite{Endres:2013sda} 
as contribution from an isolated eigenvalue (one-instanton effect). 
Note that this kind of large-order behavior would not be 
observed in supersymmetric quantities. For example, the free energy ($-\ln Z_{(1,0)}$) 
has no perturbative contribution and its trans-series expansion starts with one-instanton effect~\cite{Endres:2013sda,Nishigaki:2014ija}. 
Another example is the one-point functions of the even-power operators $\frac1N\tr\phi^{2\ell}$ or 
$\frac1N\tr B^\ell$ ($\ell\in\bm N$) 
that will be considered in the next subsection and 
appendix \ref{app:evenone-pt}. Their genus expansions terminate at some finite order. 
This fact originates from huge cancellation due to the supersymmetry. 
Thus we recognize that in order to predict nonperturbative effect from the large-order behavior 
in perturbation theory, we have to consider non-supersymmetric operators in general 
to prevent cancellation in perturbative series. Here we have explicitly observed it 
for the odd-power operators (\ref{eq:one-ptgenusexp}).

\subsection{Computation of full contribution}
\label{subsec:ACbyF}
In this subsection, we compute the full contribution to $I_{m,\,n}$ for $n\in\bm N$ 
including nonuniversal parts. 
The evaluation of the nonuniversal parts is relevant to fix the mixing terms in (\ref{eq:mixing}) and to 
obtain two-point functions in the next paper~\cite{paperII}. 

We change the integration variable as $x= -2+4t$ in \eqref{eq:Imn} 
and express $I_{m,\,n}$ in terms of the hypergeometric function: 
\begin{align}
I_{m,\,n} & = 4(4i)^mb^n\oint_{[0,1],\,t}t^{\frac{m}{2}}(1-t)^{\frac{m}{2}}\left(1-\frac{4}{b^2}t\right)^{\frac{n}{2}} 
\nn \\
& = \frac{(4i)^{m+1}}{2\pi}(1-(-1)^m)b^n\frac{\Gamma\left(\frac{m}{2}+1\right)^2}{\Gamma(m+2)}\,
F\left(-\frac{n}{2}, \frac{m}{2}+1, m+2; \frac{4}{b^2}\right). 
\label{eq:Imn3}
\end{align}
We have fixed the branch of $(x^2-4)^{\frac{m}{2}}$ as 
\be
(x^2-4)^{\frac{m}{2}} = (4e^{i\frac{\pi}{2}})^m t^{\frac{m}{2}}(1-t)^{\frac{m}{2}},
\ee
where $t^{\frac{m}{2}}(1-t)^{\frac{m}{2}}$ has no phase factor for $t=\tau+i0$ and $0<\tau<1$. 
In the first line of (\ref{eq:Imn3}), $m$ must be an integer for the integrand to be single-valued along the contour 
surrounding the cut $[0,\,1]$.  
The hypergeometric function itself in the last line of (\ref{eq:Imn3}) 
\be
F\left(-\frac{n}{2}, \frac{m}{2}+1, m+2; \frac{4}{b^2}\right)=\sum_{p=0}^\infty
\frac{\left(-\frac{n}{2}\right)_p\left(\frac{m}{2}+1\right)_p}{(m+2)_p}\,\frac{1}{p!}\left(\frac{4}{b^2}\right)^p
\ee
with $(x)_p\equiv x(x+1)\cdots(x+p-1)$ and $(x)_0\equiv 1$ 
is not well-defined for $m=-3,-5,-7,\cdots$ because of $(m+2)_p$ in the denominator. However, together with the prefactor $1/\Gamma(m+2)$, 
it becomes nonsingular:   
\be
\frac{1}{\Gamma(m+2)}\,\frac{1}{(m+2)_p} = \frac{1}{\Gamma(m+2+p)} . 
\label{eq:hypergeom}
\ee
Let us consider the case $m=-2r-1$ ($r\geq 2$) in (\ref{eq:1pt3}). 
Then, noting that nonvanishing contribution in the sum of $p$ starts with $2r$ due to (\ref{eq:hypergeom}), 
we shift the variable $p$ as 
$p \to p+2r$ and recast $I_{-2r-1,\,n}$ in terms of another hypergeometric function that is clearly nonsingular : 
\be
I_{-2r-1,\,n}=\frac{\left(-\frac{n}{2}\right)_{2r}}{(2r)!}\,b^{n-4r}F\left(2r-\frac{n}{2}, r+\frac12, 2r+1; \frac{4}{b^2}\right) .
\label{eq:Imn4}
\ee 
The expression itself is also correct for $r=0, \,1$. 
The $m=1$ case in \eqref{eq:Imn3} is also well-defined and from \eqref{eq:1pt3} 
the disk amplitude becomes 
\be
\vev{\frac{1}{N}\tr\phi^{n}}_0^{(1,0)}=-\frac12I_{1,\,n}=b^n F\left(-\frac{n}{2}, \frac32, 3;\frac{4}{b^2}\right). 
\label{eq:1pt_planar}
\ee
This agrees with the result obtained previously (eq.~(3.1) in~\cite{Kuroki:2012nt}).

When $n$ is even: $n=2\ell$ ($\ell\in\bm N$), $I_{-2r-1,\,2\ell}$ is not null for $2r\leq \ell$ due to 
the factor $\left(-\frac{n}{2}\right)_{2r}=(-\ell)_{2r}$ in (\ref{eq:Imn4}). 
In this case, $I_{-2r-1,\,2\ell}$ reduces a polynomial of $b^2=2+\mu^2$ with the degree 
$(\ell-2r)$: 
\be
I_{-2r-1,\,2\ell}= \frac{(-\ell)_{2r}}{(2r)!}b^{2\ell-4r}\sum_{p=0}^{\ell-2r}\frac{(2r-\ell)_p\left(r+\frac12\right)_p}{(2r+1)_p}
\frac{1}{p!}
\left(\frac{4}{b^2}\right)^p.
\label{eq:Imn_n:even}
\ee
Plugging this and the result for $C_{j,\,r}$ in appendix~\ref{app:recursion} into (\ref{eq:1pt3}) presents 
the full contribution to the higher-genus one-point functions 
$\vev{\frac{1}{N}\tr \phi^{2\ell}}_j$ ($j\in\bm N$) as polynomials of $b^2=2+\mu^2$. 
Since $I_{-2r-1,\,2\ell}=0$ for $\ell\leq 2r-1$ and $I_{-4j-1,\,8j}=1$ ($j\in\bm N$) 
by \eqref{eq:Imn_n:even}, it is easy to see from \eqref{eq:1pt3}
\bea
& & \vev{\frac{1}{N}\tr\phi^{2\ell}}_j^{(1,0)}=0 \qquad \mbox{for} \quad \ell\leq 4j-1,
\label{eq:1pt_even_null} \\
& & \vev{\frac{1}{N}\tr\phi^{8j}}_j^{(1,0)}= C_{j,\,2j}=\frac{(4j-1)!!}{2j+1}, 
\label{eq:1pt_8j}
\eea
where we have used (\ref{C_r=2j}). 
In addition to the planar contribution 
\be
\vev{\frac{1}{N}\tr\phi^2}^{(1,0)}_0=\mu^2, \quad \vev{\frac{1}{N}\tr\phi^4}^{(1,0)}_0=1+\mu^4,
\quad \vev{\frac{1}{N}\tr\phi^6}^{(1,0)}_0=3\mu^2+\mu^6, 
\cdots,
\ee 
the first few nonvanishing expressions at each genus of $h=1,\,2,\,3$ are 
\begin{align}
& \vev{\frac{1}{N}\tr\phi^8}^{(1,0)}_1=1,\qquad \vev{\frac{1}{N}\tr\phi^{10}}^{(1,0)}_1=5\mu^2, 
\quad \vev{\frac{1}{N}\tr\phi^{12}}^{(1,0)}_1=10+15\mu^4,
\cdots , \nn \\
& \vev{\frac{1}{N}\tr\phi^{16}}^{(1,0)}_2=21, \quad \vev{\frac{1}{N}\tr\phi^{18}}^{(1,0)}_2=189\mu^2,\quad 
\vev{\frac{1}{N}\tr\phi^{20}}_2= 483+945\mu^4, \cdots , \nn \\
& \vev{\frac{1}{N}\tr\phi^{24}}^{(1,0)}_3=1485, \quad \vev{\frac{1}{N}\tr\phi^{26}}^{(1,0)}_3=19305\mu^2,\quad 
\vev{\frac{1}{N}\tr\phi^{28}}_3= 56628+135135\mu^4,  \nn \\
& \cdots.
\label{eq:1pt_even_f}
\end{align}
On the other hand, for odd $n$: $n=2k+1$ ($k\in\bm N\cup\{0\}$), 
the full contribution to $\vev{\frac{1}{N}\tr \phi^{2k+1}}_h$ ($h\in\bm N\cup\{0\}$) 
exhibits singular behavior in the double scaling limit (\ref{eq:dsl}) due to the argument of the hypergeometric functions 
$\frac{4}{b^2}=\left(1+N^{-\frac23}\frac{s}{4}\right)^{-1}$ 
approaching $1$ from below~\footnote{This is similar to the planar case~(\ref{eq:1pt_planar}) that has been discussed 
in~\cite{Kuroki:2012nt}. The variable $\omega$ used there corresponds to $N^{-\frac23}\frac{s}{4}$.}.  
{}From (\ref{eq:1pt_planar}), we have 
\begin{align}
\vev{\frac1N\tr\phi}_0^{(1,0)}& =\frac{64}{15 \pi }+N^{-\frac23}\frac{4 s}{3 \pi }+
N^{-\frac43}\frac{s^2 \ln s }{8 \pi }+{\cal  O}\left((N^{-\frac43}\ln N)\,s^2\right), \nn \\
\vev{\frac1N\tr\phi^3}_0^{(1,0)}& =\frac{1024}{105 \pi }+N^{-\frac23}\frac{32 s}{5 \pi }
+N^{-\frac43}\frac{s^2}{\pi }
+N^{-2}\frac{s^3 \ln s}{16 \pi }+{\cal  O}\left((N^{-2}\ln N)\,s^3\right), \nn \\
\vev{\frac1N\tr\phi^5}_0^{(1,0)}& =\frac{8192}{315 \pi }+N^{-\frac23}\frac{512 s }{21 \pi }
+N^{-\frac43}\frac{8 s^2}{\pi }+N^{-2}\frac{5 s^3}{6 \pi }
+N^{-\frac83}\frac{5 s^4 \ln s}{128 \pi }\nn \\
& \hspace{5mm}+{\cal  O}\left((N^{-\frac83}\ln N) \,s^4\right), \nn \\
 \cdots &
\label{eq:one-ptatgenus0}
\end{align}
at the planar level. 
The terms carrying the factor $\ln s$ are the leading nonanalytic terms that are regarded as universal contribution. 
As mentioned in \eqref{eq:lnN}, terms of order $(N^{-\frac23(k+2)}\ln N) s^{k+2}$ 
are polynomials of $s$ and nonuniversal. 
For higher-genus cases ($h=1,2,3$), (\ref{eq:1pt3}) with (\ref{eq:Imn4}) leads to 
\begin{align}
&\vev{\frac1N\tr\phi}_1^{(1,0)}=-N^{\frac23}\frac{1}{48 \pi  s }
+{\cal  O}\left(\ln(N^{-\frac23}s)\right), \qquad 
\vev{\frac1N\tr\phi^3}_1^{(1,0)}=-\frac{\ln s}{32 \pi }
+{\cal  O}\left((\ln N)\,s\right), \nn \\
& \vev{\frac1N\tr\phi^5}_1^{(1,0)}=-\frac{1}{12 \pi }-N^{-\frac23}\frac{5 s \ln s }{64 \pi }+{\cal  O}\left((N^{-\frac23}\ln N)\,s\right), \qquad 
\cdots, 
\label{eq:one-ptatgenus1}
\end{align}
\begin{align}
& \vev{\frac1N\tr\phi}^{(1,0)}_2= -N^{\frac83}\frac{1}{192\pi s^4} +\cO(N^2s^{-3}), \qquad 
\vev{\frac1N\tr\phi^3}^{(1,0)}_2=N^2\frac{1}{384\pi s^3} + \cO(N^{\frac43}s^{-2}), \nn \\
& \vev{\frac1N\tr\phi^5}^{(1,0)}_2=-N^{\frac43}\frac{5}{1536\pi s^2} +\cO(N^{\frac23}s^{-1}), \qquad 
\cdots, 
\end{align}
\begin{align}
&\vev{\frac1N\tr\phi}^{(1,0)}_3=-N^{\frac{14}{3}}\frac{5}{288\pi s^7} +\cO(N^4s^{-6}), \qquad
\vev{\frac1N\tr\phi^3}^{(1,0)}_3=N^4\frac{5}{1152\pi s^6} +\cO(N^{\frac{10}{3}}s^{-5}), \nn \\
&\vev{\frac1N\tr\phi^5}^{(1,0)}_3=-N^{\frac{10}{3}}\frac{5}{2304\pi s^5} +\cO(N^{\frac83}s^{-4}), \qquad 
\cdots. 
\label{eq:1pt_h=3}
\end{align}
In (\ref{eq:one-ptatgenus0})-(\ref{eq:1pt_h=3}), the leading nonanalytic contribution at $s=0$ 
agrees with \eqref{eq:1pt_h1} and (\ref{eq:1pt_h2}).

\section{Instanton corrections to disk amplitudes}
\label{sec:inst}
\setcounter{equation}{0}
In this section, we compute instanton corrections 
to one-point functions at the planar level. In \cite{Endres:2013sda}, it is shown that 
isolated eigenvalues around the origin 
give rise to nonperturbative effect and trigger spontaneous supersymmetry breaking. 
In fact, the origin is a saddle point of the effective potential for isolated eigenvalues 
in the large-$N$ limit, and hence this configuration can be referred to 
as instantons~\footnote{It is well-known that such an isolated eigenvalue 
plays a role of nonperturbative effect in noncritical string theory 
\cite{David:1992za,Hanada:2004im,Kawai:2004pj,Ishibashi:2005zf,Marino:2007te,
Marino:2008vx,Schiappa:2013opa}.}.  

Here following the derivation in \cite{Hanada:2004im}, let us compute 
instanton contribution to the one-point functions in the $(1,0)$ filling fraction. 
Namely, we are interested in the one-point functions in the presence of the instanton. 
The partition function in the $(1,0)$ sector is expressed by integrals along the positive real axis 
${\bm R}_+\equiv [0,\,\infty)$ 
with respect to all $N$ eigenvalues. In the large-$N$ limit, the perturbative partition function 
without instanton contribution comes from the integral region $[a,\,b]$ for each eigenvalue. 
In the decomposition of the partition function 
\be
Z_{(1,0)}=\sum_{p=0}^N \left.Z_{(1,0)}\right|_{\text{$p$-inst.}}, 
\label{eq:Z_inst-decom}
\ee
the partition function with $p$ instantons involved is defined as $p$ eigenvalues 
integrated over the outside of $[a,\,b]$: 
\bea
\left.Z_{(1,0)}\right|_{\text{$p$-inst.}} & = & \binomi{N}{p}\tilde{C}_N\int_a^b\prod_{i=1}^{N-p}d\lambda_i
\int_{{\bm R}_+\setminus[a,b]}\prod_{j=N-p+1}^N d\lambda_j\left(\prod_{n=1}^N 2\lambda_n\right) \triangle(\lambda^2)^2 \nn \\
 & & \hspace{55mm} \times e^{-N\sum_{i=1}^N\frac12(\lambda_i-\mu^2)^2}. 
\label{eq:Z_pinst}
\eea
The expectation value $\left.\vev{\cO}^{(1,0)}\right|_{\text{$p$-inst.}}$of an operator ${\cal O}$ 
under the partition function $\left.Z_{(1,0)}\right|_{\text{$p$-inst.}}$ is defined accordingly. 
Then the expectation value of the operator $\cO$ under $Z_{(1,0)}$ can be written as
\be
\vev{\cO}^{(1,0)}
=\sum_{p=0}^N
\frac{\left.Z_{(1,0)}\right|_{\text{$p$-inst.}}}{Z_{(1,0)}}\left.\vev{\cO}^{(1,0)}
\right|_{\text{$p$-inst.}}. 
\label{eq:vev_pinst}
\ee
{}From~\cite{Endres:2013sda,Nishigaki:2014ija}, the partition functions behave as 
\be
\left.Z_{(1,0)}\right|_{\text{0-inst.}}=1,\qquad 
\left.Z_{(1,0)}\right|_{\text{$p$-inst.}}=\left(\frac{e^{-\frac43s^{3/2}}}{16\pi s^{3/2}}\right)^p\times\left[1+\cO(s^{-3/2})\right] 
\label{eq:Z_pinst_dsl}
\ee
in the double scaling limit with $s$ finite but large. 
Hence the expansion in (\ref{eq:vev_pinst}) by the instanton weight 
$e^{-\frac43s^{3/2}}/(16\pi s^{3/2})$ becomes 
\bea
\vev{\cO}^{(1,0)} & = & \left.\vev{\cO}^{(1,0)}\right|_{\text{0-inst.}} \nn \\
& & +\left.Z_{(1,0)}\right|_{\text{1-inst.}} \left(\left.\vev{\cO}^{(1,0)}\right|_{\text{1-inst.}}-\left.\vev{\cO}^{(1,0)}\right|_{\text{0-inst.}}\right) 
\nn \\
& & +\left.Z_{(1,0)}\right|_{\text{2-inst.}} \left(\left.\vev{\cO}^{(1,0)}\right|_{\text{2-inst.}}-\left.\vev{\cO}^{(1,0)}\right|_{\text{0-inst.}}\right) 
\nn \\
& & +\left(\left.Z_{(1,0)}\right|_{\text{1-inst.}}\right)^2 \left(-\left.\vev{\cO}^{(1,0)}\right|_{\text{1-inst.}}+\left.\vev{\cO}^{(1,0)}\right|_{\text{0-inst.}}\right) 
\nn \\
& & +\left.Z_{(1,0)}\right|_{\text{3-inst.}} \left(\left.\vev{\cO}^{(1,0)}\right|_{\text{3-inst.}}-\left.\vev{\cO}^{(1,0)}\right|_{\text{0-inst.}}\right) 
\nn \\
& & +\left.Z_{(1,0)}\right|_{\text{1-inst.}} \left.Z_{(1,0)}\right|_{\text{2-inst.}} 
\left(-\left.\vev{\cO}^{(1,0)}\right|_{\text{2-inst.}}-\left.\vev{\cO}^{(1,0)}\right|_{\text{1-inst.}}+2\left.\vev{\cO}^{(1,0)}\right|_{\text{0-inst.}}\right) 
\nn \\
& & +\left(\left.Z_{(1,0)}\right|_{\text{1-inst.}}\right)^3 \left(\left.\vev{\cO}^{(1,0)}\right|_{\text{1-inst.}}-\left.\vev{\cO}^{(1,0)}\right|_{\text{0-inst.}}\right) 
\nn \\
& & +(\mbox{contribution from the total instanton number $p\geq 4$}). 
\label{eq:O_exp_inst}
\eea
On the r.h.s., the third and fourth lines express contribution from the total instanton number two ($p=2$), 
and the fifth, sixth and seventh lines from three ($p=3$).  

\subsection{Schwinger-Dyson equations for $\phi^2$-resolvent at the presence of instantons}
Let us consider the case where the number of instantons is $p=\cO(N^0)\ll N$. 
Almost ($N-p$) eigenvalues belong to the support $[a, b]$ that allows the usual $1/N$ or genus 
expansion, whereas the remaining small number ($p$) of eigenvalues are outside of the support.  

For a single-trace operator $\cO$, we express the planar part (without handles, but with boundaries 
by instantons allowed) of 
$\left.\vev{\cO}^{(1,0)}\right|_{\text{$p$-inst.}}$ 
as $\left.\vev{\cO}_0^{(1,0)}\right|_{\text{$p$-inst.}}$. 
When $\left.\vev{\cO}^{(1,0)}\right|_{\text{$0$-inst.}}=\cO(N^0)$ as usual, $\left.\vev{\cO}_0^{(1,0)}\right|_{\text{$p$-inst.}}$ has 
contribution of $\cO(N^0)$ from the $(N-p)$ eigenvalues and those of $\cO(p/N)$ from the $p$ eigenvalues. 
The latter is the deviation from the usual planar contribution due to the instantons. 
For the two-point planar contribution $\vev{\cO_1\cO_2}^{(1,0)}_0$, 
the large-$N$ factorization holds even in the presence of the instantons: 
\be
\left.\vev{\cO_1\cO_2}^{(1,0)}_0\right|_{\text{$p$-inst.}}=
\left.\vev{\cO_1}_0^{(1,0)}\right|_{\text{$p$-inst.}}\left.\vev{\cO_2}_0^{(1,0)}\right|_{\text{$p$-inst.}}\times\left(1+\cO(N^{-2})\right) 
\label{eq:factorization_pinst}
\ee
for $p=\cO(N^0)$. Plugging (\ref{eq:O_exp_inst}) into the r.h.s. leads to the expansion of the two-point function 
by the instanton weight.   
The Schwinger-Dyson equation for the $\phi^2$-resolvent (\ref{eq:R2op}) derived in~\cite{Kuroki:2012nt} reads 
\be
z\vev{R_2(z)\,R_2(z)}_0^{(1,0)}=(z^2-\mu^2z)\vev{R_2(z)}_0^{(1,0)}-z+\mu^2-C_0
\label{eq:R2_SD}
\ee
with $C_0=\vev{\frac{1}{N}\tr\phi^2}^{(1,0)}_0$. 
{}From the expansion (\ref{eq:O_exp_inst}) for $\vev{R_2(z)}_0^{(1,0)}$ and $C_0$, 
and from (\ref{eq:factorization_pinst}) with $\cO_1=\cO_2=R_2(z)$, we have 
the expansion of (\ref{eq:R2_SD}) by the instanton weight: \\
\noindent  
\underline{0-instanton sector}: 
\be
z\left(\left.\vev{R_2(z)}^{(1,0)}_0\right|_{\text{0-inst.}}\right)^2=(z^2-\mu^2z) \left.\vev{R_2(z)}^{(1,0)}_0\right|_{\text{0-inst.}}-z+\mu^2
-\left.C_0\right|_{\text{0-inst.}}, 
\label{eq:SD_0inst}
\ee
\underline{1-instanton sector}: 
\begin{align}
&  z\left\{\left(\left.\vev{R_2(z)}^{(1,0)}_0\right|_{\text{1-inst.}}\right)^2-\left(\left.\vev{R_2(z)}^{(1,0)}_0\right|_{\text{0-inst.}}\right)^2\right\} \nn \\
&  = (z^2-\mu^2z)\left\{\left.\vev{R_2(z)}^{(1,0)}_0\right|_{\text{1-inst.}}-\left.\vev{R_2(z)}^{(1,0)}_0\right|_{\text{0-inst.}}\right\} 
-\left(\left.C_0\right|_{\text{1-inst.}}-\left.C_0\right|_{\text{0-inst.}}\right).
\label{eq:SD_1inst1}
\end{align}
Plugging (\ref{eq:SD_0inst}) into (\ref{eq:SD_1inst1}) simplifies the equation as 
\be
z\left(\left.\vev{R_2(z)}^{(1,0)}_0\right|_{\text{1-inst.}}\right)^2=(z^2-\mu^2z) \left.\vev{R_2(z)}^{(1,0)}_0\right|_{\text{1-inst.}}-z+\mu^2
-\left.C_0\right|_{\text{1-inst.}}. 
\label{eq:SD_1inst2}
\ee
For higher instantons, we can reduce the equations in a similar manner to obtain \\
\underline{$p$-instanton sector}: 
\be
z\left(\left.\vev{R_2(z)}^{(1,0)}_0\right|_{\text{$p$-inst.}}\right)^2=(z^2-\mu^2z) \left.\vev{R_2(z)}^{(1,0)}_0\right|_{\text{$p$-inst.}}-z+\mu^2
-\left.C_0\right|_{\text{$p$-inst.}}, 
\label{eq:SD_pinst}
\ee
which is solved by 
\be
\left.\vev{R_2(z)}_0^{(1,0)}\right|_{\text{$p$-inst.}}=
\frac12\left[z-\mu^2-\sqrt{(z-\mu^2)^2-4+\frac{4(\mu^2-\left.C_0\right|_{\text{$p$-inst.}})}{z}}\,\right]. 
\label{eq:R2_sol1}
\ee
In $p=0$ case, $C_0$ is determined by requiring no singularity other than the cut $[a^2, \,b^2]$ 
that corresponds to the perturbative saddle points. Thus 
\be
 \left.C_0\right|_{\text{0-inst.}}=\mu^2.
 \label{eq:C0_0inst}
\ee
{}From (\ref{eq:R2_sol1}), we see that the eigenvalues relevant to instantons are around the origin, i.e. 
\be
\oint_{0, \,z}\left.\vev{R_2(z)}_0^{(1,0)}\right|_{\text{$p$-inst.}}= \frac{p}{N}.
\label{eq:pinst_cond}
\ee
Note that this equation indicates that $\mu^2-\left.C_0\right|_{\text{$p$-inst.}}$ is of ${\cal O}(1/N)$ 
and hence from the factorization (\ref{eq:factorization_pinst}), 
the expression (\ref{eq:R2_sol1}) is valid within the linear order of 
$\mu^2-\left.C_0\right|_{\text{$p$-inst.}}$: 
\be
\left.\vev{R_2(z)}_0^{(1,0)}\right|_{\text{$p$-inst.}}=\left.\vev{R_2(z)}_0^{(1,0)}\right|_{\text{0-inst.}}-
\frac{\mu^2-\left.C_0\right|_{\text{$p$-inst.}}}{z\sqrt{(z-\mu^2)^2-4}}+{\cal O}((\mu^2-\left.C_0\right|_{\text{$p$-inst.}})^2). 
\label{eq:R2_sol2}
\ee
In imposing the condition (\ref{eq:pinst_cond}), we take care of the branch cut of $\sqrt{(z-\mu^2)^2-4}=\sqrt{(z-a^2)(z-b^2)}$ and see
$\left.\sqrt{(z-\mu^2)^2-4}\right|_{z=0}=-ab=-\sqrt{\mu^4-4}$. Then the solution becomes 
\begin{align}
& \left.C_0\right|_{\text{$p$-inst.}}=\mu^2-\frac{p}{N}\sqrt{\mu^4-4}+\cO((p/N)^2), 
\label{eq:C0_pinst}
\\
& \left.\vev{R_2(z)}_0^{(1,0)}\right|_{\text{$p$-inst.}}
=\left.\vev{R_2(z)}_0^{(1,0)}\right|_{\text{0-inst.}}-
\frac{p}{N}\frac{\sqrt{\mu^4-4}}{z\sqrt{(z-\mu^2)^2-4}}+{\cal O}((p/N)^2). 
\label{eq:R2_sol3}
\end{align}
$\left.C\right|_{\text{$p$-inst.}}-\mu^2=\left.\vev{\frac{1}{N}\tr(\phi^2-\mu^2)}^{(1,0)}\right|_{\text{$p$-inst.}}$ 
can also be obtained from (\ref{eq:Z_pinst_dsl}) as
\be
\left.C\right|_{\text{$p$-inst.}}-\mu^2=N^{-2}\frac{\partial}{\partial\mu^2}\ln \left.Z_{(1,0)}\right|_{\text{$p$-inst.}}
=-N^{-\frac43}2ps^{\frac12}\left(1+\cO(s^{-4/3})\right)
\ee
in the double scaling limit with $s$ finite but large, 
which is consistent with (\ref{eq:C0_pinst}) up to higher-genus contribution in the last factor $\left(1+\cO(s^{-4/3})\right)$.  

\subsection{Instanton corrections to one-point functions}
Now, instanton corrections to the disk amplitudes $\vev{\frac{1}{N}\tr\phi^n}^{(1,0)}_0$ 
($n\in\bm N$) are computed as in (\ref{eq:1pt1}):\footnote{We recall the footnote \ref{footnote:union}.} 
\be
\left.\vev{\frac{1}{N}\tr\phi^n}^{(1,0)}_0\right|_{\text{$p$-inst.}} 
= \oint_{0\cup [a,\,b],\,z}2z\cdot z^n\left.\vev{R_2(z^2)}^{(1,0)}_0\right|_{\text{$p$-inst.}}. 
\ee 
The integral encircling the origin vanishes due to the factor $2z\cdot z^n$, and we have
\be
\left.\vev{\frac{1}{N}\tr\phi^n}^{(1,0)}_0\right|_{\text{$p$-inst.}} =
\left.\vev{\frac{1}{N}\tr\phi^n}^{(1,0)}_0\right|_{\text{0-inst.}} 
+\frac{p}{N}\sqrt{\mu^4-4}\,I_{-1,\,n-2} + \cO\left((p/N)^2\right)
\label{eq:1pt_pinst}
\ee
by setting $x=\mu^2-z^2$ and \eqref{eq:Imn}. 
{}From (\ref{eq:Imn4}), 
\be
I_{-1,\,n-2}=b^{n-2}\,F\left(1-\frac{n}{2},\,\frac12,\,1;\,\frac{4}{b^2}\right).
\ee
For odd $n$, the first few expressions of $I_{-1,\,n-2}$ are given by
\bea
I_{-1,\,-1} & = & -\frac{1}{2\pi}\ln s +\cO\left((\ln N)s^0\right), \nn \\
I_{-1,\,1} & = & \frac{4}{\pi} -N^{-\frac23}\frac{1}{4\pi}\,s\ln s +\cO\left((N^{-\frac23}\ln N) s\right), \nn \\
I_{-1, \,3} & = & \frac{32}{3\pi}+N^{-\frac23}\frac{6}{\pi}\,s - N^{-\frac43}\frac{3}{16\pi}\,s^2\ln s + \cO\left((N^{-\frac43}\ln N)s^2\right), \nn \\
\cdots .& & 
\label{eq:I-1_odd}
\eea
The leading nonanalytic term of $I_{-1,\,2k-1}$ reads 
\be
I_{-1,\,2k-1}=-\left(N^{-\frac23}\right)^k\frac{1}{2\pi^{\frac32}}
\frac{\Gamma(k+\frac12)}{k!}\,s^k\ln s+ \mbox{(less singular at $s=0$)},
\label{eq:I-1_nonana}
\ee
which is consistent with (\ref{eq:Iresult1}). 
For even $n$, $I_{-1,\,n-2}$ reduces to a polynomial of $b^2(=4+N^{-\frac23}s)$: 
\bea
I_{-1,\,0} & = & 1, \nn \\
I_{-1,\,2} & = & 2+N^{-\frac23}s, \nn \\
I_{-1,\,4} & = & 6+N^{-\frac23}4s+N^{-\frac43}s^2, \nn \\
I_{-1,\,6} & = & 20+N^{-\frac23}18s+N^{-\frac43}6s^2+N^{-2}s^3, \nn \\
\cdots . & & 
\label{eq:I-1_even}
\eea  
Thus the difference of the $p$-instanton contribution from the 0-instanton one: 
\be
\Delta\left.\vev{\frac{1}{N}\tr\phi^n}^{(1,0)}_0\right|_{\text{$p$-inst.}}\equiv 
\left.\vev{\frac{1}{N}\tr\phi^n}^{(1,0)}_0\right|_{\text{$p$-inst.}}
-\left.\vev{\frac{1}{N}\tr\phi^n}^{(1,0)}_0\right|_{\text{0-inst.}}
\ee
becomes 
\begin{align}
& \Delta\left.\vev{\frac{1}{N}\tr\phi}^{(1,0)}_0\right|_{\text{$p$-inst.}}
=N^{-\frac43}p\left[-\frac{1}{\pi}\,s^{\frac12}\ln s + \cO\left((\ln N) s^{\frac12}\right)\right] +\cO\left((p/N)^2\right), \nn \\
& \Delta\left.\vev{\frac{1}{N}\tr\phi^3}^{(1,0)}_0\right|_{\text{$p$-inst.}}
=N^{-\frac43}p\left[\frac{8}{\pi}\,s^{\frac12} -N^{-\frac23}\frac{1}{2\pi}\,s^{\frac32}\ln s +\cO\left((N^{-\frac23}\ln N)s^{\frac32}\right)\right]
+\cO\left((p/N)^2\right), 
\nn \\
& \Delta\left.\vev{\frac{1}{N}\tr\phi^5}^{(1,0)}_0\right|_{\text{$p$-inst.}}
=N^{-\frac43}p\left[\frac{64}{3\pi}\,s^{\frac12} +N^{-\frac23}\frac{44}{3\pi}\,s^{\frac32} -N^{-\frac43}\frac{3}{8\pi}\,s^{\frac52}\ln s + 
\cO\left((N^{-\frac43}\ln N)s^{\frac52}\right)\right] \nn \\
& \hspace{45mm}+\cO\left((p/N)^2\right), \nn \\
& \hspace{17mm} \cdots, 
\label{eq:Delta_1ptpinst_odd}
\end{align}
and 
\begin{align}
& \Delta\left.\vev{\frac{1}{N}\tr\phi^2}^{(1,0)}_0\right|_{\text{$p$-inst.}}
=N^{-\frac43}p\left[2s^{\frac12} +N^{-\frac23}\frac14s^{\frac32} + \cO\left(N^{-\frac43}s^{\frac52}\right)\right]+\cO\left((p/N)^2\right), \nn \\
& \Delta\left.\vev{\frac{1}{N}\tr\phi^4}^{(1,0)}_0\right|_{\text{$p$-inst.}}
=N^{-\frac43}p\left[4s^{\frac12} +N^{-\frac23}\frac52s^{\frac32} + \cO\left(N^{-\frac43}s^{\frac52}\right)\right]+\cO\left((p/N)^2\right), \nn \\
& \Delta\left.\vev{\frac{1}{N}\tr\phi^6}^{(1,0)}_0\right|_{\text{$p$-inst.}}
=N^{-\frac43}p\left[12s^{\frac12} +N^{-\frac23}\frac{19}{2}s^{\frac32} + N^{-\frac43}\frac{93}{32}s^{\frac52} + \cO\left(N^{-2}s^{\frac72}\right)\right] \nn \\
& \hspace{45mm}+\cO\left((p/N)^2\right), \nn \\
& \hspace{17mm} \cdots.
\label{eq:Delta_1ptpinst_even}
\end{align}

\subsection{Operator mixing}
For odd $n$, we expect from (\ref{eq:I-1_nonana}) 
that the terms with the $\ln s$ factor would have a universal meaning in (\ref{eq:Delta_1ptpinst_odd}), 
which reads 
\be
-\left(N^{-\frac23}\right)^{k+2}\frac{p}{\pi^{\frac32}}\frac{\Gamma(k+\frac12)}{k!}\,
s^{k+\frac12}\ln s
\ee
with $n=2k+1$. 
In fact, the power of $N$ is the same as that in the perturbative result (\ref{eq:one-ptgenusexp}), 
which shows that 
nonperturbative as well as perturbative contribution equally survive in the double scaling limit as should be. 
On the other hand, polynomials of $s$ in (\ref{eq:I-1_odd}) appear as nonanalytic terms 
of half-integer powers 
due to the factor $\sqrt{\mu^4-4}=N^{-\frac13}2s^{\frac12}\left[1+\cO(N^{-\frac23}s)\right]$ 
in the one-point functions (\ref{eq:Delta_1ptpinst_odd}). 
Since their nonanalyticity is stronger than the would-be universal term of $s^{k+\frac12}\ln s$ 
at $s=0$ , 
there seems no reason to throw away such terms. 
This is in contrast to the perturbative case where we can safely discard polynomials of $s$ in (\ref{eq:one-ptatgenus0}) as  
nonuniversal parts. 

In~\cite{Kuroki:2012nt}, we have encountered similar difficulty in computing cylinder amplitudes and discussed the operator mixing to resolve it. 
The operator mixing in \cite{Kuroki:2012nt} is given as 
\begin{align}
&\Phi_1=\frac1N\tr\phi, \nn\\
&\Phi_3=\frac1N\tr\phi^3
-\frac{4}{\pi}\left(1+\bar{\alpha}_{3,2}^{(1)}\,\omega +{\cal O}(\omega^2)\right)\frac1N\tr\phi^2, \nn\\
&\Phi_5=\frac1N\tr\phi^5
-\frac{4}{\pi}\left(1+\bar{\alpha}_{5,4}^{(1)}\,\omega +{\cal O}(\omega^2)\right)\frac1N\tr\phi^4 \nn \\
& \hspace{23mm} -\frac{8}{3\pi}\left(1+3(1-\bar{\alpha}_{5,4}^{(1)})\,\omega 
+{\cal O}(\omega^2)\right)\frac1N\tr\phi^2
\label{Phi-mixing}
\end{align}
with $\omega=N^{-\frac23}\frac{s}{4}$ for the $(1,0)$ filling fraction. $\bar{\alpha}_{3,2}^{(1)}$ 
and $\bar{\alpha}_{5,4}^{(1)}$ are numerical constants 
undetermined from the cylinder amplitudes among $\Phi_1$, $\Phi_3$ and $\Phi_5$. 
Since the one-point functions of even-power operators (\ref{eq:Delta_1ptpinst_even}) have 
half-integer powers of $s$, 
it is reasonable to expect that the operator mixing cancels the half-integer powers 
between (\ref{eq:Delta_1ptpinst_odd}) and (\ref{eq:Delta_1ptpinst_even}). 
Straightforward calculations actually prove that is the case, leading to 
\begin{align}
& \Delta\left.\vev{\Phi_1}^{(1,0)}_0\right|_{\text{$p$-inst.}}=
\left(N^{-\frac23}\right)^2p\left[-\frac{1}{\pi}\,s^{\frac12}\ln s + \cO\left((\ln N) s^{\frac12}\right)\right] +\cO\left((p/N)^2\right), \nn \\
& \Delta\left.\vev{\Phi_3}^{(1,0)}_0\right|_{\text{$p$-inst.}}=
\left(N^{-\frac23}\right)^3p\left[-\frac{1}{2\pi}\,s^{\frac32}\ln s + \cO\left((\ln N) s^{\frac32}\right)\right] +\cO\left((p/N)^2\right), \nn \\
& \Delta\left.\vev{\Phi_5}^{(1,0)}_0\right|_{\text{$p$-inst.}}=
\left(N^{-\frac23}\right)^4p\left[-\frac{3}{8\pi}\,s^{\frac52}\ln s + \cO\left((\ln N) s^{\frac52}\right)\right] +\cO\left((p/N)^2\right). 
\label{eq:1pt_Phipinst}
\end{align}
In particular, $\bar{\alpha}_{3,2}^{(1)}$ and $\bar{\alpha}_{5,4}^{(1)}$ cancel 
in the leading terms in (\ref{eq:1pt_Phipinst}) and remain undetermined again.  
Similarly to the perturbative case, the last terms of $\cO\left((\ln N) s^{k+\frac12}\right)$ 
in the square brackets are less singular at $s=0$ than the first terms. 
Hence the operator mixing discussed in the cylinder amplitudes (\ref{Phi-mixing}) works 
even at the nonperturbative instanton contribution. 

Thus we confirm again that the double scaling limit keeps valid even in this case. 
By multiplying the ``wave function renormalization" factor $N^{\frac23(k+2)}$, 
(\ref{eq:1pt_Phipinst}) becomes finite in this limit~\footnote{This is also the case with the $c=0$ bosonic string theory 
as discussed in \cite{Hanada:2004im}.}. 
The one-point functions in the presence of instantons also has 
the logarithmic singularity. 
The $s$-dependence of $\Delta\left.\vev{\Phi_{2k+1}}^{(1,0)}_0\right|_{\text{$p$-inst.}}$ is $s^{k+\frac12}\ln s$, 
which is different from the planar result of the zero-instanton sector $s^{k+2}\ln s$. 
The difference of the power $s^{-\frac32}$ is proportional to $g_s$, 
and it can be interpreted as contribution from a hole created by the instanton to the one-point function
~\footnote{Note that backreaction to the instantons, i.e. influence of the presence of the operator on the instanton background, 
is not taken into account in this calculation.}. 
This result would be important in trying to identify a counterpart of the matrix model instanton 
in the type IIA side. 
In appendix \ref{app:distortion}, we present other derivation 
of \eqref{eq:R2_sol3} based on distortion of the eigenvalue distribution by instantons.

\section{Correlation functions in the full sector}
\label{sec:fullsector}
\setcounter{equation}{0}
 So far we have considered the correlation functions with a definite filling fraction, say $(\nu_+,\nu_-)=(1,0)$. 
 In this section we discuss those in the full sector, namely summed over the filling fractions. 
At first sight, it seems difficult to formulate them because of the vanishing total partition function \eqref{eq:totalZ}.  
In order to regularize it and get well-defined correlation functions, 
we introduce a factor $e^{-i\alpha\nu_-N}$ with a small parameter $\alpha$ in front of $Z_{(\nu_+,\nu_-)}$~\cite{Endres:2013sda}: 
\begin{align}
&Z_{\alpha}\equiv
\sum_{\nu_-N=0}^{N}\frac{N!}{(\nu_+N)!(\nu_-N)!}\,e^{-i\alpha\nu_-N}Z_{(\nu_+,\nu_-)}
=(1-e^{-i\alpha})^NZ_{(1,0)}. 
\label{eq:regularizedZ}
\end{align} 
Correspondingly, regularized correlation functions among $K$ single-trace operators are 
\begin{align}
\vev{\prod_{a=1}^K\frac{1}{N}\tr{\cal O}_a(\phi)}_{\alpha} & \equiv \frac{\tilde{C}_N}{Z_{\alpha}} 
\sum_{\nu_-N=0}^{N}\frac{N!}{(\nu_+N)!(\nu_-N)!}\,e^{-i\alpha\nu_-N} \nn \\
& \hspace{7mm} \times \int_0^{\infty}\left(\prod_{i=1}^{\nu_+N}2\lambda_id\lambda_i\right)
\int_{-\infty}^0\left(\prod_{j=\nu_+N+1}^N2\lambda_jd\lambda_j\right)
\triangle(\lambda^2) \nn \\
&  \hspace{17mm}\times 
\left(\prod_{a=1}^K\frac{1}{N}\sum_{i=1}^N{\cal O}_a(\lambda_i)\right)\,e^{-N\sum_{m=1}^N\frac12(\lambda_m^2-\mu^2)^2}. 
\label{eq:Kpt_alpha}
\end{align}
As discussed in \cite{Kuroki:2009yg, Kuroki:2010au}, the regularization parameter $\alpha$ 
could also be interpreted as an external field 
in discussing spontaneous supersymmetry breaking, e.g. 
the magnetic field in the spontaneous magnetization in spin systems. 
In~\cite{Endres:2013sda}, the one-point function $\vev{\frac1N\tr (iB)}_\alpha$, 
equivalently $\vev{\frac1N\tr (\phi^2-\mu^2)}_\alpha$, 
has been computed as one of the order parameters of the supersymmetry breaking. 
There, the result is independent of $\alpha$ and 
has a well-defined limit for $\alpha\to 0$. In general, when all of the operators 
$\frac{1}{N}\tr{\cal O}_a$ ($a=1,\cdots, K$) 
are even for the sign flip $\lambda_j\to -\lambda_j$ 
($j=\nu_+N+1,\cdots,N$) considered in (\ref{eq:Z_MM}), the $\alpha$-dependence 
between the numerator and the denominator cancels each other 
in (\ref{eq:Kpt_alpha}). Namely, the regularization works for correlators among even-power operators.   
On the other hand, this is not the case with odd-power operators 
$\frac1N\tr\phi^{2k+1}$ ($k\in\bm N\cup\{0\}$) involved. 
In what follows, we argue that nontrivial $\alpha$-dependence appearing 
in correlation functions of the odd-power operators 
can be absorbed into a ``wave-function renormalization'' and then 
the limit $\alpha\to 0$ can be safely taken to reduce to the correlation functions to those 
in the $(1,0)$ filling fraction.

\subsection{One-point functions}
\label{subsec:fullone-point}
The odd-power operator $\frac{1}{N}\tr\phi^{2k+1}$ changes under the sign flip as 
\be
\frac{1}{N}\sum_{i=1}^N\lambda_i^{2k+1} \to \frac{1}{N}\sum_{i=1}^{\nu_+N}\lambda_i^{2k+1} -\frac{1}{N}\sum_{j=\nu_+N+1}^{N}\lambda_j^{2k+1}
\label{eq:lambda_flip}
\ee
in terms of the eigenvalues. By using permutation symmetries with respect to 
$\lambda_1,\cdots, \lambda_N$ in the eigenvalue-integrals (
(\ref{eq:Kpt_alpha}) with $K=1$ and $\cO_1(\phi)=\phi^{2k+1}$), 
the r.h.s. of (\ref{eq:lambda_flip}) can be replaced by $(\nu_+-\nu_-)\lambda_1$ 
and further by $(\nu_+-\nu_-) \frac{1}{N}\sum_{i=1}^N\lambda_i$ in the integrals. 
Then, the one-point function $\vev{\frac{1}{N}\tr\phi^{2k+1}}_\alpha$ becomes 
\be
\vev{\frac{1}{N}\tr\phi^{2k+1}}_\alpha = \frac{1}{(1-e^{-i\alpha})^N}\left\{\sum_{\nu_-N=0}^N\frac{N!\,(\nu_+-\nu_-)}{(\nu_+N)!(\nu_-N)!}\,
(-e^{-i\alpha})^{\nu_-N}\right\}\vev{\frac{1}{N}\tr\phi^{2k+1}}^{(1,0)}.
\ee
The sum in the curly bracket on the r.h.s. is computed as 
\bea
\sum_{n=0}^N\frac{N!}{n!(N-n)!}\,\left(1-2\frac{n}{N}\right)
(-e^{-i\alpha})^n & = & \left(1-\frac{2i}{N}\partial_\alpha\right)(1-e^{-i\alpha})^N \nn \\
 & = & (1-e^{-i\alpha})^N\left\{1+\frac{2e^{-i\alpha}}{1-e^{-i\alpha}}\right\}.
\eea
Thus we find 
\be
\vev{\frac1N\tr\phi^{2k+1}}_{\alpha}= C(\alpha)\,\vev{\frac1N\tr\phi^{2k+1}}^{(1,0)}
\label{eq:1pt_full}
\ee
with 
\be
C(\alpha)\equiv -i\cot\frac{\alpha}{2}.
\label{eq:Calpha}
\ee
Although $C(\alpha)$ diverges as $\alpha\to 0$, (\ref{eq:1pt_full}) seems to suggest that the divergence could be absorbed into 
a kind of ``wave function renormalization'' 
\be
\frac1N\tr\phi^{2k+1}\to C(\alpha)^{-1} \frac1N\tr\phi^{2k+1}
\label{eq:wf_ren_alpha}
\ee
in computing correlation functions in the full sector. 
Then the result is reduced to the one in the $(1,0)$ filling fraction that is finite 
and well-defined as $\alpha\to 0$. 
Of course, we need to check whether (\ref{eq:wf_ren_alpha}) is valid or not in other cases.  
Let us consider the two-point functions of odd-power operators as the first nontrivial check.

\subsection{Two-point functions}
\label{subsec:fulltwo-point}
As in the case of the one-point functions, 
we first consider the sign flip in the product of the two odd-power operators $\frac{1}{N}\tr\phi^{2k+1}\,\frac{1}{N}\tr\phi^{2\ell+1}$. 
In terms of the eigenvalues, it changes to 
\be
 \left(\frac{1}{N}\sum_{i_1=1}^{\nu_+N}\lambda_{i_1}^{2k+1} -\frac{1}{N}\sum_{j_1=\nu_+N+1}^{N}\lambda_{j_1}^{2k+1}\right)
 \left(\frac{1}{N}\sum_{i_2=1}^{\nu_+N}\lambda_{i_2}^{2\ell+1} 
-\frac{1}{N}\sum_{j_2=\nu_+N+1}^{N}\lambda_{j_2}^{2\ell+1}\right). 
\ee
Expanding the product and extracting terms with $i_1=i_2$ or $j_1=j_2$ leads to 
\begin{align}
& \frac{1}{N^2}\sum_{i=1}^{\nu_+N}\lambda_i^{2k+1}\lambda_i^{2\ell+1}+\frac{1}{N^2}\sum_{j=\nu_+N+1}^{N}\lambda_j^{2k+1}\lambda_j^{2\ell+1}
+\frac{1}{N^2}\sum_{i_1\neq i_2}\lambda_{i_1}^{2k+1}\lambda_{i_2}^{2\ell+1}
+\frac{1}{N^2}\sum_{j_1\neq j_2}\lambda_{j_1}^{2k+1}\lambda_{j_2}^{2\ell+1} 
\nn \\
&  -\frac{1}{N^2}\sum_{i=1}^{\nu_+N}\sum_{j=\nu_+N+1}^N\left(\lambda_i^{2k+1}\lambda_j^{2\ell+1}+\lambda_j^{2k+1}\lambda_i^{2\ell+1}\right),
\end{align}
where the sum of $i_1$ and $i_2$ ($j_1$ and $j_2$) is understood to run from 1 to $\nu_+N$ (from $\nu_+N+1$ to $N$). 
Use of the permutation symmetry of the eigenvalue-integrals allows us to replace this by~\footnote{This manipulation is similar to 
the one presented in appendix B of~\cite{Kuroki:2012nt}.} 
\be
(\nu_+-\nu_-)^2\lambda_1^{2k+1}\lambda_2^{2\ell+1}+
\frac{1}{2N}\left(\lambda_1^{2k+1}-\lambda_2^{2k+1}\right)\left(\lambda_1^{2k+1}-\lambda_2^{2\ell+1}\right)
\ee
in the integrals. Then we have 
\begin{align}
& \vev{\frac{1}{N}\tr\phi^{2k+1}\,\frac{1}{N}\tr\phi^{2\ell+1}}_\alpha = \frac{1}{(1-e^{-i\alpha})^N}\sum_{\nu_-N=0}^N
\frac{N!}{(\nu_+N)!(\nu_-N)!}\, (-e^{-i\alpha})^{\nu_-N} \nn \\
& \hspace{7mm} \times \left[(\nu_+-\nu_-)^2\vev{\lambda_1^{2k+1}\lambda_2^{2\ell+1}}^{(1,0)} 
+\frac{1}{2N}\vev{\left(\lambda_1^{2k+1}-\lambda_2^{2k+1}\right)\left(\lambda_1^{2k+1}-\lambda_2^{2\ell+1}\right)}_C^{(1,0)}\right]. 
\end{align}
Note that 
\be
\vev{\frac{1}{N}\tr\phi^{2k+1}\,\frac{1}{N}\tr\phi^{2\ell+1}}^{(1,0)} = \vev{\lambda_1^{2k+1}\lambda_2^{2\ell+1}}^{(1,0)} 
+ \frac{1}{2N}\vev{\left(\lambda_1^{2k+1}-\lambda_2^{2k+1}\right)\left(\lambda_1^{2k+1}-\lambda_2^{2\ell+1}\right)}_C^{(1,0)}
\ee
and the second term on the r.h.s. is negligible compared to the first term in the double scaling limit. 
Thus under the prescription in taking the limits: 
\begin{enumerate}
\item
take the double scaling limit first,
\item
then, turn off $\alpha$, 
\end{enumerate}
we obtain 
\bea
\vev{\frac{1}{N}\tr\phi^{2k+1}\,\frac{1}{N}\tr\phi^{2\ell+1}}_\alpha
& = & \frac{1}{(1-e^{-i\alpha})^N}\left\{\sum_{n=0}^N\frac{N!}{n!(N-n)!}\left(1-2\frac{n}{N}\right)^2(-e^{-i\alpha})^n\right\} \nn \\
& & \times \vev{\frac{1}{N}\tr\phi^{2k+1}\,\frac{1}{N}\tr\phi^{2\ell+1}}^{(1,0)}. 
\eea
After computing the sum, extracting the connected pieces from this expression leads to 
\begin{align}
\vev{\frac{1}{N}\tr\phi^{2k+1}\,\frac{1}{N}\tr\phi^{2\ell+1}}_{\alpha,\,C} 
 &=C(\alpha)^2\vev{\frac{1}{N}\tr\phi^{2k+1}\,\frac{1}{N}\tr\phi^{2\ell+1}}_C^{(1,0)} \nn \\
 & \hspace{6mm}-\frac{1}{N}(C(\alpha)^2-1)\vev{\frac{1}{N}\tr\phi^{2k+1}\,\frac{1}{N}\tr\phi^{2\ell+1}}^{(1,0)}.
\label{eq:2pt_alpha} 
\end{align} 
The two-point function in the last term consists of the connected and disconnected pieces: 
 $\vev{\frac{1}{N}\tr\phi^{2k+1}\,\frac{1}{N}\tr\phi^{2\ell+1}}_C^{(1,0)}$ and 
$\vev{\frac{1}{N}\tr\phi^{2k+1}}^{(1,0)}\vev{\frac{1}{N}\tr\phi^{2\ell+1}}^{(1,0)}$.  
They are of the same order in the double scaling limit as far as the universal parts are concerned. 
Namely, we assume that in the double scaling limit, non-universal parts which would become dominant in each correlation function 
are subtracted in advance.  
Then the last term on the r.h.s. of (\ref{eq:2pt_alpha}) can be neglected in the double scaling limit due to the prefactor $\frac{1}{N}$. 
In conclusion, we arrive at 
\be
\left.\vev{\frac{1}{N}\tr\phi^{2k+1}\,\frac{1}{N}\tr\phi^{2\ell+1}}_{\alpha,\,C} \right|_{\rm univ.}
 =C(\alpha)^2\left.\vev{\frac{1}{N}\tr\phi^{2k+1}\,\frac{1}{N}\tr\phi^{2\ell+1}}_C^{(1,0)}\right|_{\rm univ.} ,
\label{eq:2pt_alpha_f}
\ee
concerning the universal parts in the double scaling limit. 
Thus we again find that the two-point functions of the ``renormalized" operators 
$C(\alpha)^{-1}\frac1N\tr\phi^{2k+1}$ are independent of $\alpha$ and reduced to those 
in the $(\nu_+,\nu_-)=(1,0)$ sector in the $\alpha\rightarrow 0$ limit. 

In summary, the one- and two-point correlation functions of ``renormalized" odd-power operators
\begin{align}
\widehat{\Phi}_k=\frac{1}{C(\alpha)}N^{\frac23(k+2)}
\left(\frac{1}{N}\tr\phi^{2k+1}-\text{(nonuniversal parts)}\right)
\end{align}
are all finite in the prescription of the limit where we take first the double scaling limit, then $\alpha\rightarrow 0$ limit. 
In the above equation, ``(nonuniversal parts)" indicates both of the mixing terms mentioned in \eqref{eq:mixing} 
and nonuniversal parts of $\frac1N\tr\phi^{2k+1}$ itself. They may be more dominant 
than the universal part in the double scaling limit unless we subtract it in advance. 
Thus the argument in this section validates concentrating on correlation functions in the $(1,0)$ sector.

\section{Discussions}
\label{sec:discussion}
In this paper, we have computed one-point functions of the operators 
$\frac{1}{N}\tr\phi^n$ ($n\in\bm N$) to all order of genus expansion 
and their instanton contribution in the supersymmetric double-well matrix model, 
which extends the work of correlation functions at the planar level~\cite{Kuroki:2012nt}. 
The matrix model is proposed to describe two-dimensional type IIA superstring theory 
on a nontrivial Ramond-Ramond background~\cite{Kuroki:2013qpa}.  
The operators with even $n$ (even-power operators) are protected by supersymmetry, 
while those with odd $n$ (odd-power operators) are not. 
We have seen that this difference is reflected by qualitatively different behavior in the correlation functions. 
For example, genus-expansion of the even $n$ case terminates at some order, 
whereas the odd $n$ case yields non-Borel summable series. 
The divergence is due to the coefficients of the series growing as $(2h)!$ for a large genus $h$, 
which has been recognized as a characteristic feature of string perturbation series~\cite{Shenker:1990uf}. 
This indicates that operators unprotected by supersymmetry play an essential role 
to understand superstring theory from the corresponding matrix model. 
{}From the non-Borel summable asymptotic series, we can read nonperturbative ambiguity 
that turns out to be of the same order as instanton effects found in~\cite{Endres:2013sda,Nishigaki:2014ija}. 
The idea of resurgence suggests that the ambiguity from the perturbative series is canceled 
with one arising from fluctuations around the instanton background 
(for example, see~\cite{Schiappa:2013opa,Pasquetti:2009jg,Aniceto:2011nu,Dunne:2012ae}). 
It is intriguing to check whether it works as well in our matrix model or superstring theory with its target 
supersymmetry spontaneously broken~\cite{paperIII}.    
 
It is discussed in~\cite{Kuroki:2012nt,Kuroki:2013qpa} that 
single-trace operators with operator mixing in our matrix model corresponds 
to integrated vertex operators in the type IIA superstring theory. 
The explicit form of the operator mixing is presented there based 
on the result of planar two-point (cylinder) amplitudes. 
We have seen here that the operator mixing is also consistent with nonperturbative instanton contribution 
to the one-point functions. 
In addition, the difference of the filling fraction $(\nu_+-\nu_-)$ in the matrix model is proportional 
to the strength of the Ramond-Ramond background flux~\cite{Kuroki:2012nt,Kuroki:2013qpa}. 
Although the correlation functions have been computed at a fixed sector of the filling fraction, 
typically the $(1,0)$ sector here, we has shown that the computation 
by the total partition function summed over the filling fractions is regularized 
by a ``wave-function renormalization'' factor and yields the same result as in the $(1,0)$ sector. 
It would be interesting to consider the meaning of the regularization in the type IIA superstring side, 
which may give new insight to the structure of vacua in the superstring theory. 

In the next papers~\cite{paperII,paperIII}, we will present the computation of two-point functions 
in the matrix model to all order in genus expansion, and discuss the further consistency 
of the operator mixing and resurgence.

\section*{Acknowledgements}
We would like to thank Satoshi~Iso, Hirohiko~Shimada, Shinobu~Hikami and Shinsuke~M.~Nishigaki 
for useful discussions and comments. 
The work of T.~K. is supported in part by a Grant-in-Aid for Scientific Research (C), 25400274, 16K05335. 
The work of F.~S. is supported in part by a Grant-in-Aid for Scientific Research (C), 25400289. 
We would like to thank the Yukawa Institute for Theoretical Physics at Kyoto University 
for hospitality during the workshop YITP-W-15-12 
"Developments in String Theory and Quantum Field Theory," 
and YITP-W-16-05 "Strings and Fields 2016," 
where part of this work was carried out.

\appendix
\section{Solution of the recursion relation (\ref{eq:recursion})}
\label{app:recursion}
\setcounter{equation}{0}
In this appendix, we present a solution of the recursion relation (\ref{eq:recursion}). 

\subsection{$C_{j,\,3j-1}$, $C_{j,\,2j}$}
In the case $r=3j+2$, by noting (\ref{eq:Cbdy}), the recursion relation (\ref{eq:recursion}) 
is reduced to 
\be
\frac{1}{(6j+3)!!}C_{j+1,\,3j+2}=\frac23\frac{1}{j+1}\frac{1}{(6j-3)!!}C_{j,\,3j-1}.
\ee
In terms of $D_j\equiv \frac{1}{(6j-3)!!}C_{j,\,3j-1}$, it is easy to solve this as 
\be
D_{j+1}=  \frac23\frac{1}{j+1}D_j 
= \cdots =  \left(\frac23\right)^j\frac{1}{(j+1)!}D_1.
\ee
{}From $D_1=\frac{1}{3!!}C_{1,2}=\frac13$, we have 
\be
D_j=\frac12\left(\frac23\right)^j\frac{1}{j!},  
\ee 
and thus 
\be
C_{j,\,3j-1}= \frac12\left(\frac23\right)^j\frac{(6j-3)!!}{j!}  \qquad(j\in\bm N). 
\label{C_r=3j-1}
\ee

In the case of $r=2j+2$, we can similarly obtain 
\begin{align}
\frac{C_{j+1,\,2j+2}}{(4j+3)!!}=\frac{2j+1}{2j+3}\frac{C_{j,\,2j}}{(4j-1)!!}
\end{align}
which leads to a solution: 
\be
C_{j,\,2j}=\frac{(4j-1)!!}{2j+1}   \qquad(j\in\bm N). 
\label{C_r=2j}
\ee

\subsection{$C_{j,\,2j+1},\,C_{j,\,2j+2},\,C_{j,\,2j+3}$}
For $r=2j+3$, the recursion relation (\ref{eq:recursion}) becomes 
\be
\frac{j+2}{(4j+5)!!}C_{j+1,\,2j+3}=\frac{j+1}{(4j+1)!!}C_{j,\,2j+1}+\frac{1}{2j+1}, 
\ee
where we have used (\ref{C_r=2j}). 
By considering 
$E_j\equiv \frac{j+1}{(4j+1)!!}C_{j,\,2j+1}$, we have 
\be
E_j = \sum_{\ell=2}^j\frac{1}{2\ell-1}.
\ee
Therefore, 
\bea
& & C_{j,\,2j+1} = \frac{(4j+1)!!}{j+1}\sum_{\ell=2}^j\frac{1}{2\ell-1}  \qquad (j\geq 2), 
\nn\\
& & C_{1,\,3} =  0. 
\label{C_r=2j+1}
\eea

Repeating a similar procedure for $r=2j+4, \,2j+5$, we obtain 
\bea
& & C_{j,\,2j+2} = 2\frac{(4j+3)!!}{2j+3}\sum_{\ell'=2}^{j-1}\frac{1}{\ell'+1}
\sum_{\ell=2}^{\ell'}\frac{1}{2\ell-1}\qquad (j\geq 3), \nn \\
& & C_{1,\,4} = C_{2,\,6}=0,
\label{C_r=2j+2}
\eea
and 
\bea
& & C_{j,\,2j+3} = 2\frac{(4j+5)!!}{j+2}\sum_{\ell''=2}^{j-2}\frac{1}{2\ell''+5}
\sum_{\ell'=2}^{\ell''}\frac{1}{\ell'+1}
\sum_{\ell=2}^{\ell'}\frac{1}{2\ell-1} \qquad (j\geq 4), \nn \\
& & C_{1,\,5}=C_{2,\,7}=C_{3,\,9}=0. 
\label{C_r=2j+3}
\eea

\subsection{$C_{j,\,r}$ for general $r$}
{}From the expressions of (\ref{C_r=2j}), (\ref{C_r=2j+1}), (\ref{C_r=2j+2}), and (\ref{C_r=2j+3}), we can find out the form of $C_{j,\,r}$ 
for general $r$: 
\bea
C_{j,\,2j+r} & = & \frac12\frac{(4j+2r-1)!!}{j+\frac{r+1}{2}}\sum_{\ell_r=2}^{j-r+1}\frac{1}{\ell_r+\frac{3r-4}{2}}
\sum_{\ell_{r-1}=2}^{\ell_r}\frac{1}{\ell_{r-1}+\frac{3r-7}{2}} \times\cdots \nn \\
& & \hspace{27mm}\times \sum_{\ell_2=2}^{\ell_3}\frac{1}{\ell_2+1}\sum_{\ell_1=2}^{\ell_2}\frac{1}{\ell_1-\frac12} \qquad (1\leq r\leq j-1), 
\nn \\
C_{j,\,2j} & = & \frac12\frac{(4j-1)!!}{j+\frac12}, 
\label{Cjr}
\eea
and all the others vanish. 

In fact, when $r=j-1$ in (\ref{Cjr}), each of the $\ell_i$ ($i=1,2,\cdots, j-1$) 
appearing in the sum takes the value 2 alone, and the expression reproduces (\ref{C_r=3j-1}). 

Finally, we explicitly present the first several nonvanishing expressions for $C_{j,\,r}$: 
\bea
& & C_{1,\,2}=1,\qquad C_{2,\,4}=21,\qquad C_{2,\,5}=105, \nn \\
& & C_{3,\,6}=1485,\qquad C_{3,\,7}=18018,\qquad C_{3,\,8}=50050,
\eea
which agree with the result given in \cite{HT_2012}.

\section{Other derivation of one-point functions of even-power operators}
\label{app:evenone-pt}
\setcounter{equation}{0}
In this appendix, we compute one-point functions of the even-power operators $\frac1N\tr\phi^{2\ell}$ or $\frac{1}{N}\tr B^\ell$ ($\ell\in\bm N$) 
at arbitrary genus in a different manner from the text. 
Since these are independent of the sector of the filling fraction as discussed in~\cite{Kuroki:2012nt}, 
let us focus on the 
$(1,0)$ filling fraction case. 
By diagonalizing $\phi$ as $\phi=U\Lambda U^{\dagger}$ with $\Lambda=\text{diag}(\lambda_1,\cdots,\lambda_N)$, the partition function can be written as  
\begin{align}
Z_{(1,0)}
=&\tilde{C}_N\int d^{N^2}B\int_0^\infty\prod_{i=1}^N(d\lambda_iW''(\lambda_i))
\prod_{i>j}(W'(\lambda_i)-W'(\lambda_j))^2\,
e^{-N\tr\left(\frac12B^2+iBW'(\Lambda)\right)}
\label{eq:Zapp}
\end{align}
with $W'(x) = x^2-\mu^2$. 
Note that the following argument is valid in a more general superpotential 
as long as $Z_{(1,0)}$ does not vanish. For example, when $W'(x)$ is a polynomial of the odd degree, 
the total partition function remains nonzero. 
The argument below (\ref{eq:Zpert}) is nonperturbatively correct for that case, 
with the replacement of $\left.Z_{(1,0)}\right|_{\text{pert.}}$, 
$\left.\vev{\frac1N\tr B^\ell}^{(1,0)}\right|_{\text{pert.}}$ and $\left.\vev{\frac1N\tr W'(\phi)^\ell}^{(1,0)}\right|_{\text{pert.}}$ 
by $Z$, $\vev{\frac1N\tr B^\ell}$ and $\vev{\frac1N\tr W'(\phi)^\ell}$, respectively. 
The Nicolai mapping 
\begin{align}
h_i=W'(\lambda_i) \quad \text{or} \quad H=W'(\Lambda),
\label{eq:Nicolai}
\end{align}
recasts (\ref{eq:Zapp}) as 
\be
Z_{(1,0)}=\tilde{C}_N\int d^{N^2}B\int_{-\mu^2}^\infty\left(\prod_{i=1}^Ndh_i\right)\triangle(h)^2\,
e^{-N\tr\left(\frac12B^2+iBH\right)} .
\label{GaussianZ}
\ee
{}From \cite{Endres:2013sda,Nishigaki:2014ija},  
the effect of the lower bound of the integration region $[-\mu^2,\infty)$ with respect to $h_i$ 
is considered to be nonperturbative in the $1/N$ expansion. 
We can replace the integrals by those over the whole real axis as far as the genus expansion is concerned. 
Thus the system we will consider is reduced to the standard Gaussian matrix model: 
\begin{align}
\left.Z_{(1,0)}\right|_{\text{pert.}}
=\int d^{N^2}B\int d^{N^2}H\,e^{-N\tr\left(\frac12B^2+iBH\right)}=1.  
\label{eq:Zpert}
\end{align}
The last equality follows from the normalization (\ref{eq:measure1}).  

\subsection{$\vev{\frac1N\tr B^\ell}^{(1,0)}$} 
\label{app:B}
It is easy to see that the one-point functions 
\begin{align}
\left.\vev{\frac1N\tr B^\ell}^{(1,0)}\right|_{\text{pert.}}
=\frac{1}{\left.Z_{(1,0)}\right|_{\text{pert.}}}
\int d^{N^2}B\int d^{N^2}H\left(\frac1N\tr B^\ell\right)e^{-N\tr\left(\frac12B^2+iBH\right)}
\end{align}
($\ell\in\bm N$) vanish due to the delta function with respect to the matrix $B$ 
which arises from the $H$ integral. 
Hence 
\begin{align}
\left.\vev{\frac1N\tr B^\ell}^{(1,0)}\right|_{\text{pert.}}=0
\end{align}
in all order in the $1/N$-expansion.  

\subsection{$\vev{\frac1N\tr W'(\phi)^\ell}^{(1,0)}$}
\label{app:phieven}
Via the mapping \eqref{eq:Nicolai}, the one-point functions 
$\left.\vev{\frac1N\tr W'(\phi)^\ell}^{(1,0)}\right|_{\text{pert.}}$ ($\ell\in\bm N$) becomes 
\begin{align}
\left.\vev{\frac1N\tr W'(\phi)^\ell}^{(1,0)}\right|_{\text{pert.}}
=&\frac{1}{\left.Z_{(1,0)}\right|_{\text{pert.}}}\int d^{N^2}H
\left(\frac1N\tr H^\ell\right)e^{-N\tr\frac12H^2}
\nn \\
= & \tilde{C}_N \int \left(\prod_{i=1}^Ndh_i\right) \triangle(h)^2\,\left(\frac{1}{N}\sum_{i=1}^Nh_i^\ell\right)\,e^{-N\sum_{i=1}^N h_i^2}.
\label{eq:Wvev}
\end{align}
We calculate this by using the orthogonal polynomials~\footnote
{Similar calculation is found in the correlation function of two ``Wilson loops" in the one-matrix model in 
\cite{Kawamoto:2008gp}.}. These are monic given by the Hermite polynomials:  
\be
P^{(H)}_n(x) =\frac{1}{(2N)^{n/2}}\,H_n\left(\sqrt{\frac{N}{2}}\,x\right) \qquad 
(n\in\bm N\cup\{0\})
\label{eq:PHn}
\ee 
with   
\be
H_n(x) \equiv (-1)^n\,e^{x^2}\frac{d^n}{dx^n}e^{-x^2} =(-1)^nH_n(-x). 
\ee
The orthogonality 
\be
\int_{-\infty}^\infty dx\,e^{-\frac{N}{2}x^2}\,P^{(H)}_n(x)P^{(H)}_m(x) = h^{(H)}_n\delta_{n,\,m}, \qquad 
h^{(H)}_n=\sqrt{2\pi}\frac{n!}{N^{n+\frac12}}
\ee
is satisfied. 
By use of these properties and the fact that the constant $\tilde{C}_N$ is expressed as 
$\tilde{C}_N = \left(N!\prod_{k=0}^{N-1}h^{(H)}_k\right)^{-1}$,  
(\ref{eq:Wvev}) becomes 
\be
\left.\vev{\frac1N\tr W'(\phi)^\ell}^{(1,0)}\right|_{\text{pert.}}= 
\frac{1}{N}\sum_{m=0}^{N-1}\frac{1}{h^{(H)}_m}\int^\infty_{-\infty}dx \,P^{(H)}_m(x)^2\,x^\ell e^{-\frac{N}{2}x^2}. 
\label{eq:Wvev2}
\ee 
Clearly this vanishes for odd $\ell$. Let us consider the case of even $\ell$ ($\ell=2p$) in what follows. 
The orthogonal polynomials can also be expressed as 
\be
P^{(H)}_n(x) = N^{-\frac{n}{2}}\left.\partial_t^n\,e^{t\sqrt{N}\,x-\frac{t^2}{2}} \right|_{t=0}
\label{eq:PHn2}
\ee
from properties of the Hermite polynomials. After plugging this into (\ref{eq:Wvev2}), 
straightforward calculation leads to 
\be
\left.\vev{\frac1N\tr W'(\phi)^{2p}}^{(1,0)}\right|_{\text{pert.}}
= N^{-p-1}(2p)!\sum_{r=0}^p\frac{1}{2^rr!((p-r)!)^2}
\sum_{m=p-r}^{N-1}\frac{m!}{(m-p+r)!}. 
\ee 
Using the identity $\sum_{m=q}^n\binomi{m}{q}=\binomi{n+1}{q+1}$ and setting 
$n=p-r$, we finally obtain 
\bea
\left.\vev{\frac1N\tr W'(\phi)^{2p}}^{(1,0)}\right|_{\text{pert.}} &= & 
\frac{(2p-1)!!}{N^p}\,F(1-N,-p,2;2) \nn \\
& = & \frac{(2p-1)!!}{N^p}\sum_{n=0}^{\infty}\frac{(1-N)_n(-p)_n}{(2)_n}\frac{2^n}{n!}
\label{eq:Wvev3}
\eea
with $(x)\equiv x(x+1)\cdots(x+n-1)$ and $(x)_0\equiv 1$. 
Hence the sum on $n$ is actually a finite one. The first few results are explicitly given by 
\begin{align*}
&\left.\vev{\frac1N\tr W'(\phi)^2}^{(1,0)}\right|_{\text{pert.}}=1, 
&&\left.\vev{\frac1N\tr W'(\phi)^4}^{(1,0)}\right|_{\text{pert.}}=2+\frac{1}{N^2}, \\
&\left.\vev{\frac1N\tr W'(\phi)^6}^{(1,0)}\right|_{\text{pert.}}=5+\frac{10}{N^2}, 
&&\left.\vev{\frac1N\tr W'(\phi)^8}^{(1,0)}\right|_{\text{pert.}}=14+\frac{70}{N^2}+\frac{21}{N^4}.  
\end{align*}
We can check that these are consistent with (\ref{eq:1pt_even_null})-(\ref{eq:1pt_even_f}) for $W'(x) = x^2-\mu^2$.

\section{Other derivation of instanton effects}
\label{app:distortion}
\setcounter{equation}{0}
In this appendix, we reproduce the instanton effect in section~\ref{sec:inst} from the viewpoint of distortion 
of the eigenvalue distribution by the instantons following the argument given in \cite{Kawamoto:2008gp}. 
The partition function $Z_{(1,0)}$ can be written as 
\be
Z_{(1,0)}= \tilde C_N\int\left(\prod_{i=1}^N2\lambda_id\lambda_i\right)
\triangle(\lambda^2)^2\,e^{-N\sum_{i=1}^N\frac12(\lambda_i^2-\mu^2)^2}
=\tilde C_N\int\left(\prod_{i=1}^N2\lambda_id\lambda_i\right)e^{-V_{\text{eff}}}
\ee
with the effective potential 
\bea
V_{\text{eff}} &\equiv &N\sum_{i=1}^N\frac12(\lambda_i^2-\mu^2)^2-\frac12\sum_{i\neq j}\log(\lambda_i^2-\lambda_j^2)^2 \nn \\
& =&N^2\int dx\,\rho(x)\frac12(x^2-\mu^2)^2-\frac{N^2}{2}\dashint dxdy\,\rho(x)\rho(y)\log(x^2-y^2)^2 \nn \\
& &+C\left(\int dx\,\rho(x)-1\right). 
\label{eq:Veff}
\eea
$\rho(x)=\frac1N\tr\delta(x-\phi)=\frac1N\sum_{i=1}^N\delta(x-\lambda_i)$  
is the eigenvalue distribution, and $C$ is a Lagrange multiplier imposing the constraint $\int dx\,\rho(x)=1$. 
Similarly to the setting in the text, let us consider the $p$-instanton sector with $p=\cO(N^0) \ll N$, where $p$ eigenvalues are apart 
from the other $N-p$ eigenvalues. 
By relabeling the eigenvalues, it is natural to 
decompose $\rho(x)$ as 
\bea
& & \rho(x) =\rho^{(0)}(x)+\frac1N\rho^{(1)}(x), \label{eq:decomp} \\
& &\rho^{(0)}(x)=\frac1N\sum_{i=1}^{N-p}\delta(x-\lambda_i), \qquad \rho^{(1)}(x)=\sum_{i=N-p+1}^N\delta(x-\lambda_i). 
\eea
Here $\rho^{(1)}(x)$ describes distribution of the isolated $p$ eigenvalues. It follows from this definition 
that 
\begin{align}
\int dx\,\rho^{(0)}(x)=1-\frac pN, \qquad \int dx\,\rho^{(1)}(x)=p. 
\label{eq:rhoint}
\end{align}
Substituting \eqref{eq:decomp} for \eqref{eq:Veff}, we obtain 
\bea
V_{\text{eff}} & =&N^2\int dx\,\rho^{(0)}(x)\frac12(x^2-\mu^2)^2
-\frac{N^2}{2}\dashint dxdy\,\rho^{(0)}(x)\rho^{(0)}(y)\log(x^2-y^2)^2 \nn \\
& &+N\int dx\,\rho^{(1)}(x)\frac12(x^2-\mu^2)^2
-N\dashint dxdy\,\rho^{(0)}(x)\rho^{(1)}(y)\log(x^2-y^2)^2 \nn \\
& &+C\left(\int dx\,\rho^{(0)}(x)+\frac1N\int dx\,\rho^{(1)}(x)-1\right)+{\cal O}(N^0).
\label{eq:Veff2}
\eea
The saddle point equation for $\rho^{(0)}(x)$ reads 
\begin{align}
0=(x^2-\mu^2)x-\dashint dy\,\rho^{(0)}(y)\left(\frac{1}{x-y}+\frac{1}{x+y}\right)
-\frac1N\dashint dy\,\rho^{(1)}(y)\left(\frac{1}{x-y}+\frac{1}{x+y}\right)
\label{eq:SPE1}
\end{align}
for $x$ inside the support of $\rho^{(0)}(x)$. This equation implies that its solution 
also has the $1/N$-expansion 
\begin{align}
\rho^{(0)}(x)=\rho^{(0,0)}(x)+\frac1N\rho^{(0,1)}(x)+\cdots, 
\label{eq:distortion}
\end{align}
where 
$\frac1N\rho^{(0,1)}(x)$
represents distortion of the eigenvalue distribution 
$\rho^{(0,0)}(x)$ in the large-$N$ limit due to the presence of the $p$ instantons. 
The solution to the equation \eqref{eq:SPE1} in the large-$N$ limit (without the second term 
on the r.h.s.) has been already given in 
\cite{Kuroki:2009yg} as 
\begin{align}
\rho^{(0,0)}(x)=\frac{x}{\pi}\sqrt{(x^2-a^2)(b^2-x^2)} \quad \mbox{with} \quad a=\sqrt{\mu^2-2}, \quad b=\sqrt{\mu^2+2} 
\end{align}
for $x\in[a,\,b]$~\footnote{
For a general filling fraction $(\nu_+,\,\nu_-)$, it becomes 
\be
\rho^{(0,0)}(x)=\begin{cases} \frac{\nu_+}{\pi}\,x\sqrt{(x^2-a^2)(b^2-x^2)}\qquad (x\in[a,\,b]) \nn \\
                                        \frac{\nu_-}{\pi}\,|x|\sqrt{(x^2-a^2)(b^2-x^2)}\qquad (x\in[-b,\,-a]) .
                                        \end{cases}
\ee                                     
}.   
Plugging \eqref{eq:distortion} into \eqref{eq:SPE1} and {(\ref{eq:rhoint}) provides conditions on $\rho^{(0,1)}(x)$: 
\bea
0=\dashint dy\,
\rho^{(0,1)}(y)\left(\frac{1}{x-y}+\frac{1}{x+y}\right)
+\dashint dy\,
\rho^{(1)}(y)\left(\frac{1}{x-y}+\frac{1}{x+y}\right)
\label{eq:SPE2}
\eea
and 
\be
\int dx\,\rho^{(0,1)}(x)= -p.
\label{eq:rho01_int}
\ee

\subsection{$\rho^{(1)}(x)$}
In order to find $\rho^{(1)}(x)$, 
we assume that the $p$ eigenvalues are located at a saddle point $x=x_*$ outside the support of 
the perturbative configurations of a general filling fraction, i.e. $\Omega\equiv [-b,\,-a]\cup [a,\,b]$ 
and make an ansatz~\footnote{Since we are considering the case $p\ll N$, 
force between $p$ eigenvalues can be neglected in this order.}
\begin{align}
\rho^{(1)}(x)=p\,\delta(x-x_*). 
\end{align}
Then the effective potential in \eqref{eq:Veff2} becomes up to ${\cal O}(N)$ 
\begin{align}
V_{\text{eff}}=&(\mbox{$x_*$-independent part})
+Np\left(\frac12(x_*^2-\mu^2)^2-\dashint dx\,\rho^{(0)}(x)\log(x^2-x_*^2)^2\right),
\end{align}
whose saddle point equation $\partial_{x_*}V_{\text{eff}}=0$ for $x_*$ yields 
\begin{align}
0=2x_*\left(x_*^2-\mu^2-2{\text{Re}}\left.\vev{R_2(x_*^2)}_0^{(1,0)}\right|_{\text{0-inst.}}\right)  . 
\end{align}
By noting (\ref{eq:R2_sol1}) with (\ref{eq:C0_0inst}), 
the solutions are 
\begin{align}
x_*=0 \quad \mbox{or} \quad x_*\in\Omega,  
\end{align}
where the first one is appropriate to describe the position of the instantons. Thus 
\be
\rho^{(1)}(x)=p\,\delta(x). 
\label{eq:rho1f}
\ee
\subsection{$\rho^{(0,1)}(x)$}
Substituting back (\ref{eq:rho1f}) for \eqref{eq:SPE2} leads to 
\begin{align}
0=\dashint dy\,
\rho^{(0,1)}(y) \left(\frac{1}{x-y}+\frac{1}{x+y}\right)+\frac{2p}{x}.
\label{eq:SPE3}
\end{align}
In order to solve this, we introduce a complex function 
\begin{align}
G(z)\equiv\int_a^bdy 
\frac{\rho^{(0,1)}(y)}{z-y},
\end{align}
and assume that $\rho^{(0,1)}(y)$ has the same support $[a,b]$ 
as that of $\rho^{(0,0)}(y)$. 
This assumption is plausible
because the distortion $\rho^{(0,1)}$ is subleading in the $1/N$-expansion of $\rho^{(0)}$ 
in \eqref{eq:distortion}
and the support will not move by $1/N$ corrections. 
Then \eqref{eq:SPE3} becomes
\begin{align}
0=G(x)-G(-x)+\frac{2p}{x} \quad \mbox{for} \quad x\in[a,b],
\end{align}
and (\ref{eq:rho01_int}) leads to 
\begin{align}
G(z) \rightarrow -\frac{p}{z} \quad (z\rightarrow\infty).
\end{align}
Therefore, 
\be
G_-(z)\equiv\frac12\left(G(z)-G(-z)\right) = z\int_a^bdy\,\frac{\rho^{(0,1)}(y)}{z^2-y^2}
\label{eq:G-}
\ee 
satisfies following conditions:
\begin{enumerate}
\item \label{cond:1}$G_-(z)$: odd, analytic in $z\in {\bf C} \setminus \Omega$.
\item $G_-(x)\in {\bf R}$ for $x\in{\bf R} \setminus \Omega$.
\item \label{cond:3}$G_-(z)\rightarrow -\frac pz+{\cal O}(1/z^3)$ as $z\rightarrow\infty$.
\item \label{cond:4} 
$
G_-(x\pm i0)=-\frac px\mp\frac{i\pi}{2}\rho^{(0,1)}(x)$
for $x\in [a,\,b]$.
\end{enumerate}
{}From these conditions we can set  
\begin{align}
G_-(z)=-\frac pz+\frac{f(z)}{\sqrt{(z^2-a^2)(z^2-b^2)}}
\label{eq:Gansatz}
\end{align}
with $f(z)$ being odd.  From the condition 3, $f(z) = \frac{\beta}{z} + \cO(z^{-3})$ as $z\to \infty$. The analyticity at the origin in 
the condition 1 requires $f(z)= \frac{\beta}{z}$ with $\beta=-pab=-p\sqrt{\mu^4-4}$. 
Hence we arrive at  
\begin{align}
G_-(z)=-\frac pz-\frac{p\sqrt{\mu^4-4}}{z\sqrt{(z^2-a^2)(z^2-b^2)}}. 
\label{eq:G-f}
\end{align}
Comparing this with the condition \ref{cond:4}, we find the distortion 
\begin{align}
\rho^{(0,1)}(x)=\begin{cases} \frac{2}{\pi}\frac{p\sqrt{\mu^4-4}}{x\sqrt{(x^2-a^2)(b^2-x^2)}} \quad &\mbox{for}\quad x\in [a,\,b] \nn \\
0 \quad &\mbox{otherwise}.\end{cases}
\label{eq:rho01_f}
\end{align}

\subsection{Final result}
Plugging the above results into 
\be
\left.\vev{R_2(z^2)}^{(1,0)}_0\right|_{\text{$p$-inst.}} = \int dy \frac{\rho(y)}{z^2-y^2}
\ee
with (\ref{eq:decomp}) and (\ref{eq:distortion}), we have 
\begin{align}
\left.\vev{R_2(z^2)}^{(1,0)}_0\right|_{\text{$p$-inst.}} & = \left.\vev{R_2(z^2)}^{(1,0)}_0\right|_{\text{0-inst.}} 
+\frac{1}{N}\left(\frac{1}{z}G_-(z)+\frac{p}{z^2}\right) + \cO(N^{-2}) \nn \\
& = \left.\vev{R_2(z^2)}^{(1,0)}_0\right|_{\text{0-inst.}} 
-\frac{1}{N}\frac{p\sqrt{\mu^4-4}}{z^2\sqrt{(z^2-a^2)(z^2-b^2)}} + \cO(N^{-2}). 
\end{align}
It is easy to see that this is equivalent with (\ref{eq:R2_sol3}).


\end{document}